\documentclass[singlecolumn,prl,superscriptaddress,12pt]{revtex4-1}

\usepackage{soul}

\usepackage{setspace, graphicx, color, amssymb, amsmath, subfigure, verbatim, ulem,cancel}
\usepackage[colorlinks]{hyperref}
\usepackage{enumitem}
\usepackage[all]{xy}
\pdfoutput=1

	\def \e {{\epsilon}}
	\def \a {{\alpha}}
	\def \g {{\gamma}}

	\def \D {{\Delta}}
	\def \d {{\delta}}

	\def \l {{\lambda}}

	\def \s {{\sigma}}
	
	\def \b {{\beta}}

	\def \n {{\eta}}
	\def \la {{\langle}}
	\def \ra {{\rangle}}

	\def \r{{\rho}}
	
	\def \ra {{\rangle}}
	\def \la {{\langle}}

\newcommand{\beq}{\begin{equation}}
\newcommand{\eeq}{\end{equation}}
\newcommand{\beqr}{\begin{eqnarray}}
\newcommand{\eeqr}{\end{eqnarray}}
\newcommand{\beqrn}{\begin{eqnarray*}}
\newcommand{\eeqrn}{\end{eqnarray*}}
\newcommand{\beqn}{\begin{equation*}}
\newcommand{\eeqn}{\end{equation*}}
\newcommand{\bei}{\begin{itemize}}
\newcommand{\beii}{\begin{itemize} \item}
\newcommand{\eei}{\end{itemize}}
\newcommand{\bmei}{\begin{itemize} \compactlist}
\newcommand{\emei}{\end{itemize}}
\newcommand{\ben}{\begin{enumerate}}
\newcommand{\een}{\end{enumerate}}
\newcommand{\bes}{\begin{small}}
\newcommand{\ees}{\end{small}}
\newcommand{\bec}{\begin{center}}
\newcommand{\eec}{\end{center}}

\newcounter{kfoo}

\begin{document}

\begin{abstract}
Experiments show that spike-triggered stimulation performed with Bidirectional Brain-Computer-Interfaces (BBCI) can artificially strengthen connections between separate neural sites in motor cortex (MC).
What are the neuronal mechanisms responsible for these changes and how does targeted stimulation by a BBCI shape population-level synaptic connectivity? 
The present work describes a recurrent neural network model with probabilistic spiking mechanisms and plastic synapses capable of capturing both neural and synaptic activity statistics relevant to BBCI conditioning protocols. 
When spikes from a neuron recorded at one MC site trigger stimuli at a second target site after a fixed delay, the connections between sites are strengthened for spike-stimulus delays consistent with experimentally derived spike time dependent plasticity (STDP) rules. 
However, the relationship between STDP mechanisms at the level of networks, and their modification with neural implants remains poorly understood. 
Using our model, we successfully reproduces key experimental results and use analytical derivations, along with novel experimental data. We then derive optimal operational regimes for BBCIs, and formulate predictions concerning the efficacy of spike-triggered stimulation in different regimes of cortical activity.
\end{abstract}

\title{Correlation-based model of artificially induced plasticity in motor cortex by a bidirectional brain-computer interface}

\author{Guillaume Lajoie}
\affiliation{University of Washington Institute for Neuroengineering, Seattle, Washington, USA}

\author{Nedialko~I.~Krouchev}
\affiliation{Montreal Neurological Institute, McGill University, Montreal, Quebec, Canada}

\author{John~F.~Kalaska}
\affiliation{Groupe de recherche sur le syst\`eme nerveux central, D\'epartement de physiologie,
Universit\'e de Montreal, Montreal, Quebec, Canada}

\author{Adrienne~L.~Fairhall}
\affiliation{University of Washington Institute for Neuroengineering, Seattle, Washington, USA}
\affiliation{University of Washington, Dept.~of Physiology and Biophysics, Seattle, Washington, USA}
\affiliation{University of Washington, Dept.~of Physics, Seattle, Washington, USA}

\author{Eberhard E. Fetz}
\affiliation{University of Washington Institute for Neuroengineering, Seattle, Washington, USA}
\affiliation{University of Washington, Dept.~of Physiology and Biophysics, Seattle, Washington, USA}

\maketitle

%

\section{Introduction}

The cerebral cortex contains interacting neurons that form new functional connections through repeated activation patterns. 
For example, motor and somatosensory cortices are typically organized into somatotopic regions in which localized neural populations are associated with muscles or receptive fields and show varied levels of correlated activity (e.g.~\cite{Fetz1980,Fetz:1984ul,Schieber:2001jr,Mountcastle:1959ud,Jackson:2003tx}). 
Functional relationships between such neural populations are known to change over time and reinforce relevant pathways~\citep{Monfils:2005cw,Sanes:2000ug,Buonomano:1998kk,Markram:2015dq}. 
These changes are the result of plasticity mechanisms acting on myriad synaptic connections between cortical neurons. Most of them are relatively weak but can potentiate under the right conditions.
However, it is not always clear what such conditions might be, or how one can interact with them for experimental or clinical purposes.
Unanswered questions include the way local synaptic plasticity rules lead to stable, emergent functional connections, and the role of neural activity |and its statistics| in shaping such connections.
While recent and ongoing work elucidates various plasticity mechanisms at the level of individual synapses, it is still unknown how these combine to shape the recurrently connected circuits that support population-level neural computations. 

\bigskip
Bidirectional brain-computer interfaces (BBCI) capable of closed-loop recording and stimulation have enabled targeted conditioning experiments that probe these issues.  
In a seminal experiment~\cite{Jackson:2006kb}, a BBCI recorded action potentials of a neuron at one MC site (labeled $N_{\rm rec}$) and delivered spike-triggered stimuli at another ($N_{\rm stim}$) for prolonged times in freely behaving macaque monkeys.
This conditioning was able to increase functional connectivity from $N_{\rm rec}$ to $N_{\rm stim}$ (c.f.~\cite{Rebesco:2010bu}), as measured by electromyogram (EMG) of muscle muscle activation evoked by intracortical microstimulation (ICMS) in MC. Importantly, the relative strength of induced changes showed a strong dependence on the spike-stimulus delay, consistent with experimentally derived excitatory spike-time-dependent plasticity (STDP) time windows~\citep{Bi:2001uy,Caporale:2008fm,Markram:1997hd}. 
The effects of this protocol were apparent after about a day of ongoing conditioning, and lasted for several days afterwards.
Similar spike-triggered stimulation showed corticospinal connections could increase or decrease, depending on whether the postsynaptic cells were stimulated after or before arrival of the presynaptic impulses~\citep{Nishimura:2013ju}.
This BBCI protocol has potential clinical uses for recovery after injuries and scientific utility for probing information processing in neural circuits.

\bigskip
The observations outlined above suggest that STDP is involved in shaping neural connections by normal activity during free behavior, and is the central mechanism behind the success of spike-triggered conditioning. However, this could not be verified directly as current experiments only measure functional relationships between cortical sites.
Furthermore, interactions between BBCI signals and neural activity in recurrent networks are still poorly understood, and it remains unclear how BBCI protocols can be scaled up, made more efficient, and optimized for different experimental paradigms. 
For example, during spike-triggered stimulation, the spikes from a single unit are used to trigger stimulation of a neural population. While STDP can explain how the directly targeted synapses may be affected (i.e.~from the origin of the recorded spikes to the stimulated population), the observed functional changes must rely on a broader scope of plastic changes involving other neurons that are functionally related to the recorded ones. What are the relevant network mechanisms that govern these population-level changes? How can a BBCI make use of population activity to trigger optimal stimulation?

Here we advance a modeling platform capable of capturing the effects of BBCI on recurrent, plastic neuronal networks. 
Our goal is to use the simplest dynamical assumptions in a ``bottom-up" approach, motivated by neuronal and synaptic physiology, that enable the reproduction of key experimental findings from~\cite{Jackson:2006kb} at the functional level in MC and related muscles. In turn, we seek to use this model to provide insights into plasticity mechanisms in recurrent MC circuits (and other cortical regions) that are not readily accessible experimentally, as well as establish a theoretical framework upon which future BBCI protocols can be developed.

\bigskip
We build on a well-established body of work that enables analytical estimates of synaptic changes based on network statistics~\citep{Kempter:1999tn,Burkitt:2007ek,Gilson:2010eh,Gilson:2009gu,Gilson:2009iy,Gilson:2009ci,Gilson:2009eq} and compare theoretical results with experiments and numerical simulations of a probabilistic spiking network. 
In our model, every neuron is excitatory | the modulatory role of inhibition in MC is instead represented implicitly by non-homogeneous probabilistic activation rates. 
While inhibition likely plays an important role in cortical dynamics, we consider results from our exclusive use of excitation to be a significant finding, suggesting that a few key mechanisms can account for a wide range of experimental results. 
Using data from previous work as well as from novel experiments, we calibrate STDP synaptic dynamics and activity correlation timescales to those typically found in MC neural populations. 
The result is a spiking model with multiplicative excitatory STDP and stable connectivity dynamics which can reproduce three key experimental findings:
\begin{enumerate}[label=\alph*.]
\item Using a single scaling parameter, we capture both (different) timescales of onset and dissipation of conditioning effects.
\item With a simple filtered readout of our model's output, we reproduce conditioning effects as measured by muscle EMG recordings evoked by ICMS.
\item We reproduce the overall conditioning dependence on spike-triggered stimulation delay.
\end{enumerate}
Furthermore, we make the following novel findings:
\begin{enumerate}
\item The distinct temporal statistics that characterize different regimes of cortical activity have an important impact on the efficacy of BBCI protocols.
\item Multi-synaptic mechanisms can lead to changes in circuit connectivity that are not predicted by STDP of synapses directly targeted by conditioning protocols.
\item When stimulating a subset of a given neural population, we find that the overall efficacy of conditioning depends supra-linearly on the proportion of this subset.
\end{enumerate}
Together, these results outline quantifiable experimental predictions. They combine into a theoretical framework that is easily scalable and serves as a potential testbed for next-generation applications of BBCIs. We discuss ways to make use of this framework in state-dependent conditioning protocols.

\section{Results}

Our model's key assumptions are: 
(i) we only consider interactions between excitatory neurons; we revisit inhibition in the discussion ; 
(ii) neurons in MC are sparsely and randomly connected by synapses that are plastic and follow a single multiplicative STDP rule ; (iii) the spiking activity of each neuron is modeled as a probabilistic process modulated by two net inputs: prescribed ``external commands" |representing functional afferents to MC| and other neurons in the network via synaptic interactions.
 \begin{figure*}[h!]
\includegraphics{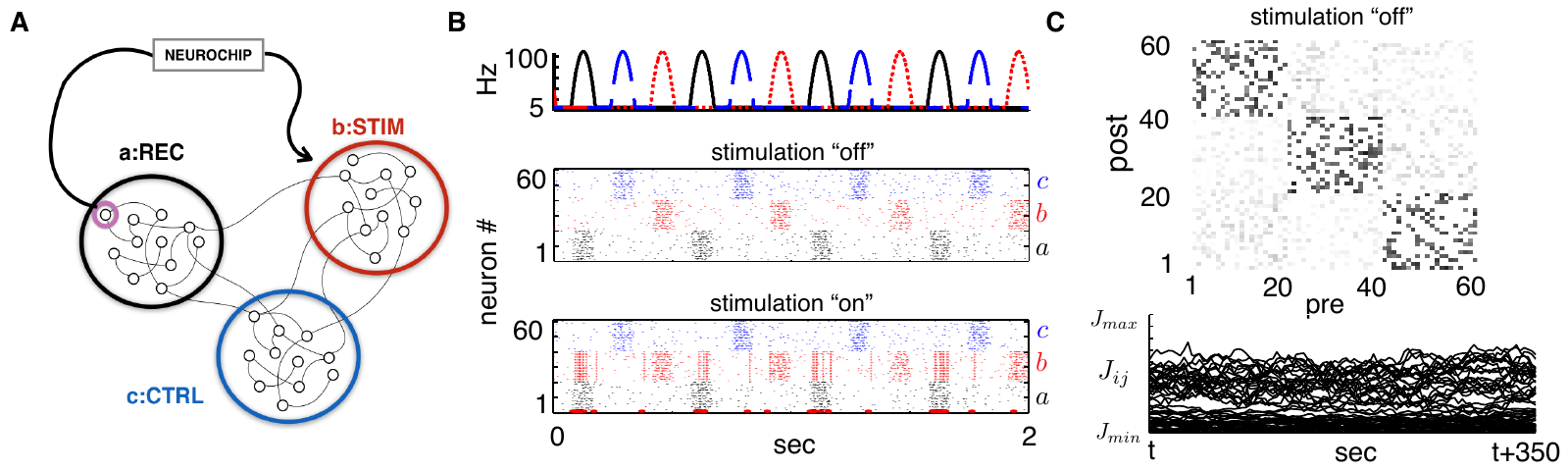}
\caption{ 
({\bf A}) Diagram of network with three functional groups ($a$, $b$, $c$). In the stimulation protocol ($\dag$), the ``neurochip" records spikes from a single neuron in group $a$ and delivers a large stimulus to group $b$ after a delay $d^\dag$. 
({\bf B}) Top: external rates $\nu_a(t)$ (solid black), $\nu_b(t)$ (dotted red) and $\nu_c(t)$ (dashed blue) delivered to each neural group. Middle: sample spike raster plot from a simulation subject to rates plotted above with sparse, homogeneous connectivity $J(0)$ ($N=60$). Bottom: same as middle but with spike-triggered stimulation ($\dag$) turned ``on". Spike from neuron number 1, circled in red, trigger the stimulation of all of population $b$ (in red) after a delay of $d^\dag=10$ ms.
({\bf C}) Top: snapshot of connectivity matrix $J(t)$ for network in B after spiking simulation of 5000 seconds (grayscale: white $=J_{min}$, black $=J_{max}$). Bottom: example time traces of individual synapses from the same simulation.
}
\label{fintro}
\end{figure*}

\medskip
We use a standard probability-based network model in which the instantaneous spiking probability rate $\l_i(t)$ of neuron $i$ changes in time and depends linearly on an external command $\nu_i(t)$, as well as the exponentially filtered spikes of other neurons in the network. Synaptic weights modulate the latter inputs and are stored in the connectivity matrix $J(t)$ which is allowed to evolve in time according to a multiplicative STDP rule $W(\D t, J_{ij})$ (see {\it Methods}). Here, $\D t$ is the inter-spike-interval between a pre-synaptic spike from neuron $j$ and a post synaptic spike from neuron $i$. 
Connectivity is chosen to be sparse and is initially drawn at random between the $N$ neurons in the network with no new synapses allowed to emerge. 
Both axonal and dendritic spike transmission delays are sampled randomly for each synapse from two respective distributions with means $\bar d^a$ and $\bar d^d$.
For numerical simulations, each neuron is treated as an independent Poisson process with rate $\l_i(t)$. 
We refer the reader to {\it Methods} for more details about the model and its implementation.

\medskip
To model the BBCI conditioning experiments,  we divide our network of MC neurons into three groups of equal sizes as illustrated in Figure~\ref{fintro} A. Neurons in each group $\a=a,b,c$ 
receive external inputs that have a group-specific modulating rate $\nu_\a(t)$ but are otherwise independent from neuron to neuron
(except for synaptic interactions).
We drive the network with idealized input rates in all numerical simulations.
These are composed of truncated sinusoidal functions offset with an equal phase difference between each group (Figure~\ref{fintro} B), and produce mean firing rates on the order of 10 Hz for most connectivity matrices. We later introduce more biologically relevant statistics for activation rates to match simple model attributes to experimental data.
To mimic the spike-triggered stimulation experiments in~\cite{Jackson:2006kb}, we assign group $a$ to be the ``Recording" site, group $b$ to be the ``Stimulation" site and $c$ to be the ``Control" site. 
Neuron $i=1$ from group $a$ is the ``recorded" neuron, whose spikes trigger stimulation of every neuron in group $b$ after a stimulation delay $d^\dag$.
Stimulation of subsets of $b$ is also considered.

\medskip
 The relationship between spiking and synaptic activity in similar models of plastic networks has been extensively studied and predictions of synaptic dynamics under a variety of contexts were analytically derived (see e.g.~\cite{Morrison:2007hp,Gutig:2003vy,Meffin:2006tg,Song:2000uc,Song:2001ws,Burkitt:2007ek,VanRossum:2000we,Rubin:2005ku,Kempter:1999tn,Gilson:2010eh,Gilson:2009gu,Gilson:2009iy,Gilson:2009ci,Gilson:2009eq}).
We build on these results and contribute novel findings that enable representations of
artificial connections from BBCIs in generalized activity regimes. Our approach largely follows that of~\cite{Gilson:2010eh} but contains key innovations which are described in detail in {\it Methods}.

\medskip
For illustration, Figure~\ref{fintro} C shows a snapshot of a connectivity matrix for a network of size $N=60$ driven by the rates shown in Figure~\ref{fintro} B and spike-triggered stimulation turned ``off", as well as traces of synaptic weights $J_{ij}(t)$ evolving in time. It is evident that some structure emerges due to external commands, and that individual synapses show ongoing fluctuations surrounding that structure.
Below, we often make use of averaged synaptic matrices $\bar J(t)$: $3\times 3$ matrices that contain group-averaged synaptic weights. In {\it Methods}, we derive analytical expressions for them, allowing us to forego lengthy spiking simulations. These expressions make use of group-averaged cross-correlation functions of different quantities that we often use throughout: for the external command rates ($\hat C(u)$), for the network spiking activity without ($C(u)$) and with spike-triggered stimulation ($C^\dag(u)$).

\subsection{Emergent synaptic structure and impact of spike-triggered stimulation}

\begin{figure*}[h!]
\includegraphics{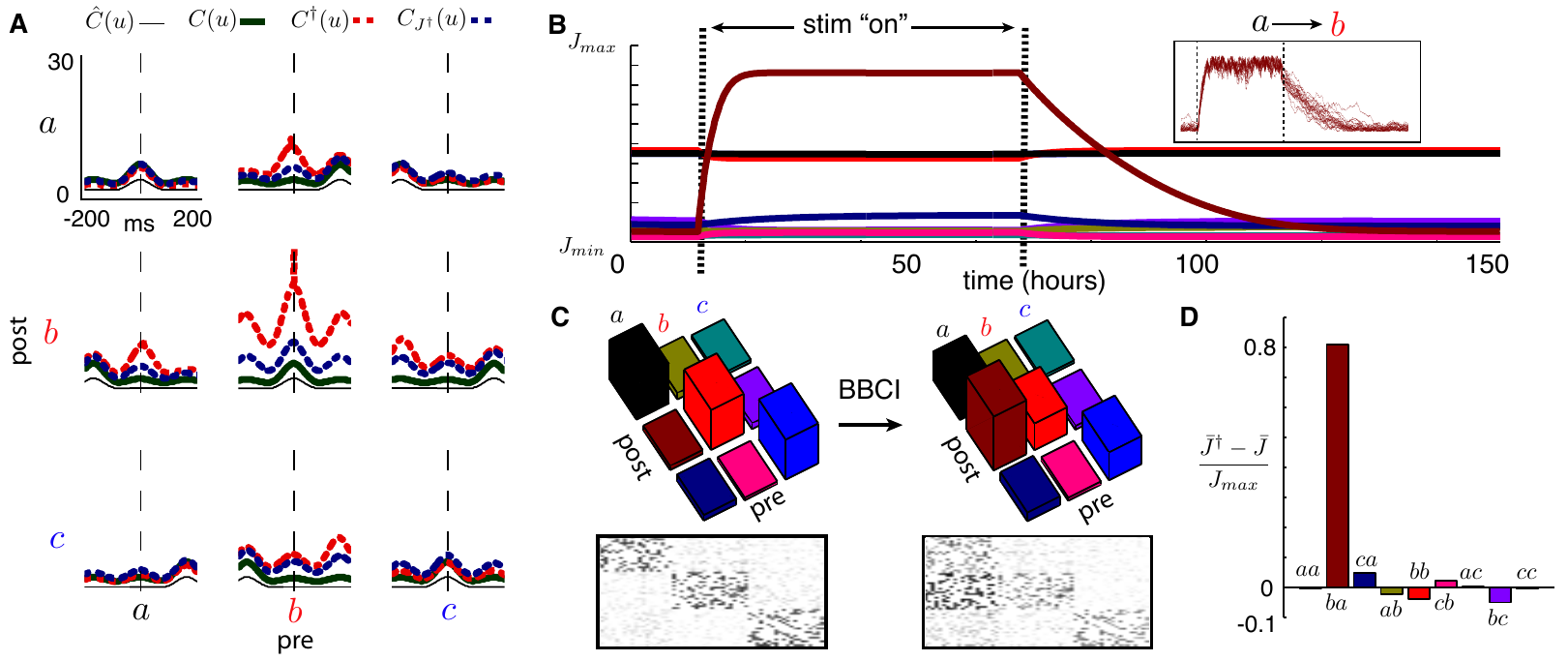}
\caption{
({\bf A}) Mean network correlations for synaptic equilibria under normal activity $C(u)$ (thick green line) and under spike-triggered stimulation $C^\dag(u)$ (dashed red line). Mean correlations for normal network activity with synaptic equilibrium obtained from stimulation ($C_{J^\dag}(u)$) is shown in dashed blue. External correlations $\hat{C}(u)$ also shown (thin black line). 
({\bf B}) Evolution of synaptic weight averages $\bar J_{\a\b}(t)$ over time, computed with analytical estimates, color-coded as in B. External rates as in Figure~\ref{fintro} B. Initiated at equilibrium (see Figure~\ref{fDS} in {\it Methods}), spike triggered stimulation ($\dag$) is switched ``on" for the indicated period. Inset shows evolutions of 10 randomly chosen synapses from group $a$ to group $b$, from corresponding spiking network simulation. 
({\bf C}) Top: plots of equilibrium matrices $\bar{J}^*$ in normal activity and under spike-triggered stimulation (left/right resp.). Bottom: snapshot of full simulated matrix $J(t)$ once equilibria are reached. 
({\bf D}) Plot of relative synaptic changes between equilibria: $(\bar{J}_{\a\b}^\dag-\bar{J}_{\a\b})/J_{max}$ with $\a,\b\in\{a,b,c\}$. 
}
\label{fbci}
\end{figure*}

When neurons in the network's three groups $a$, $b$, $c$ are subject to external commands $\nu(t)=$($\nu_a(t)$, $\nu_b(t)$, $\nu_c(t)$) with stationary statistics, their averaged connectivity evolves toward an equilibrium that reflects these inputs' correlations (c.f.~\citep{Gutig:2003vy,Meffin:2006tg}), although individual synapses may continue to fluctuate. 
This was previously observed in a number of theoretical studies (see e.g.~\cite{Morrison:2007hp,Gilson:2010eh}) and is consistent with the formation and dissociation of muscle assemblies in MC due to complex movements that are regularly performed~\citep{Buonomano:1998kk}.
The mean synaptic equilibrium $\bar{J}^*$ strongly depends on the external inputs $\nu(t)$'s correlation structure $\hat{C}(u)$ (see Figure~\ref{fbci} A). 
Indeed, a narrow peak near the origin for correlations within groups, as is the case for the periodic external rates shown in Figure~\ref{fintro} B, along with the absence of such peaks for cross-group correlations, contribute to strengthening synapses within groups and weakening those across groups. Under such conditions, what will be the impact of spike-triggered stimulation?

Figure~\ref{fbci} B shows the evolution of synaptic averages $\bar J_{\a\b}(t)$, analytically computed (see Figure~\ref{fDS} C in {\it Methods}) for a system initiated at the synaptic equilibrium associated with external rates $\nu(t)$ from Figure~\ref{fintro} B. The inset of Figure~\ref{fbci} B shows the evolution of individual synapses from group $a$ to group $b$ from full network simulations.
At 15 hours, the spike-triggered stimulation protocol is turned ``on", with a set delay $d^\dag=20$ milliseconds, and synapses start changing. In $\sim$10 hours they reach a new equilibrium which differs from the initial one in a few striking ways, as seen in Figure~\ref{fbci} C and D, where normalized differences $(\bar{J}^\dag-\bar{J})/J_{max}$ are plotted for all pre- and post-group combinations.
First, as expected and in accordance with experiments~\citep{Jackson:2006kb}, the mean strength of synapses from group $a$ ($N_{rec}$) to group $b$ ($N_{stim}$) are considerably strengthened (by about $80\%$). 
 As described in more detail below, this massive potentiation relies on two ingredients: correlated activity within group $a$ and an appropriate choice of stimulation delay $d^\dag$.

Perhaps more surprising are collateral changes in other synapses although they are of lesser magnitude. While this was previously unreported, it is consistent with unpublished data from the original spike-triggered stimulation experiment~\citep{Jackson:2006kb}. It is unclear how many of these changes are due to the particular external rate statistics and other model parameters; we return to this question below when realistic activity statistics are considered.

\medskip
We also show that spike-triggered stimulation induces novel correlation structures due to synaptic changes as illustrated in Figure~\ref{fbci} A, which plots the correlations: $\hat{C}(u)$, $C(u)$, $C^\dag(u)$ and $C_{J^\dag}(u)$. 
The latter denotes the correlations one observes under baseline activity (i.e. without ongoing stimulation) but with the equilibrium connectivity obtained after spike-triggered stimulation ($\bar{J}^{\dag*}$), i.e., at the end of spike-triggered stimulation. It is clear that every interaction involving group $b$ is considerably changed, most strongly with group $a$, including the neuron used to trigger stimulation. More surprising is the increased cross-correlations of group $b$ with itself, even though connectivity within group $b$ is not explicitly potentiated by conditioning. In fact, it is slightly depressed (Figure~\ref{fbci} D). This happens because connections from group $a$ to group $b$ are considerably enhanced, which causes the mean firing rate of group $b$ to grow and its correlations to increase. Later, we explore similar collateral changes that occur because of multi-synaptic interactions. 
We see in the next section how these correlations translate into functional changes in the network.

\medskip
Finally, Figure~\ref{fbci} B shows a crucial feature of our model: there are two timescales involved in the convergence from the normal to the artificial equilibrium, and the decay back to normal equilibrium after the end of spike-triggered stimulation. In the conditioning experiments from~\citep{Jackson:2006kb,Nishimura:2013ju}, the effect of spike-triggered stimulation was seen after about 5-24 hours of conditioning, while the changes decayed after 1 to 7 days. With the simplified external drives producing a reasonable mean firing rate of about 10Hz for individual cells, an STDP learning rate of $\n=10^{-8}$ was adequate to capture the two timescales of synaptic changes.

Thus, the simple excitatory STDP-based mechanisms in our model produce distinct
timescales for increases and decay of the strength of synaptic connections
produced by spike-triggered conditioning, in agreement with experimental observations.
This does not contradict the findings from human
psychophysical studies that feedback error-driven motor adaptation may
involve two or more different and independent parallel processes with
different temporal dynamics for learning and decay of the motor skill~\cite{Smith:2006ir}. Nevertheless, the cellular mechanisms in our model may have
some relation to the different timescales proposed to underlie motor
adaptation at the system level~\cite{Smith:2006ir}. Such relationships could be further investigated by direct experimentation and appropriate simulations.

\medskip
In summary, our model satisfies the first experimental observation from~\cite{Jackson:2006kb} we set out to reproduce (point a.~in {\it Introduction}). Indeed, we find that {\it two distinct timescales of synaptic changes (during and after conditioning) are an emergent property of our model, and tuning a single parameter is sufficient to reproduce the rates observed in experiments.}

\subsection{Simulating ICMS protocols: recovering functional changes}

Changes in correlations due to spike-triggered stimulation indicate that there is an activity-relevant effect of induced synaptic changes, which is measurable from spiking statistics (see Figure~\ref{fbci} A). We now show how this is directly observable in evoked activity patterns that are consistent with ICMS protocols employed in experiments.
In~\cite{Jackson:2006kb}, connectivity changes were inferred using intra-cortical microstimulation (ICMS) and electromyogram (EMG) recordings of the monkey's wrist muscles, as well as evoked isometric torques. To summarize, a train of ICMS stimuli lasting 50 ms was delivered to each MC site; simultaneously, EMG activity in three target muscles were recorded. The average EMG responses for repeated trials were documented for each of three MC sites (i.e. group $a$, $b$ and $c$) before and after spike-triggered conditioning.

The experiment showed that prior to conditioning, ICMS stimulation of any MC site elicited well-resolved average EMG responses, largest in one muscle but not the two others. After conditioning, ICMS stimulation of the recording site ($N_{rec}$, group $a$) not only elicited an EMG response in its associated muscle, but also in that of stimulated site ($N_{stim}$, group $b$). 
While it was conjectured that synaptic changes in MC were responsible for the changes, this could not be verified directly.
 \begin{figure*}[h!]
\includegraphics{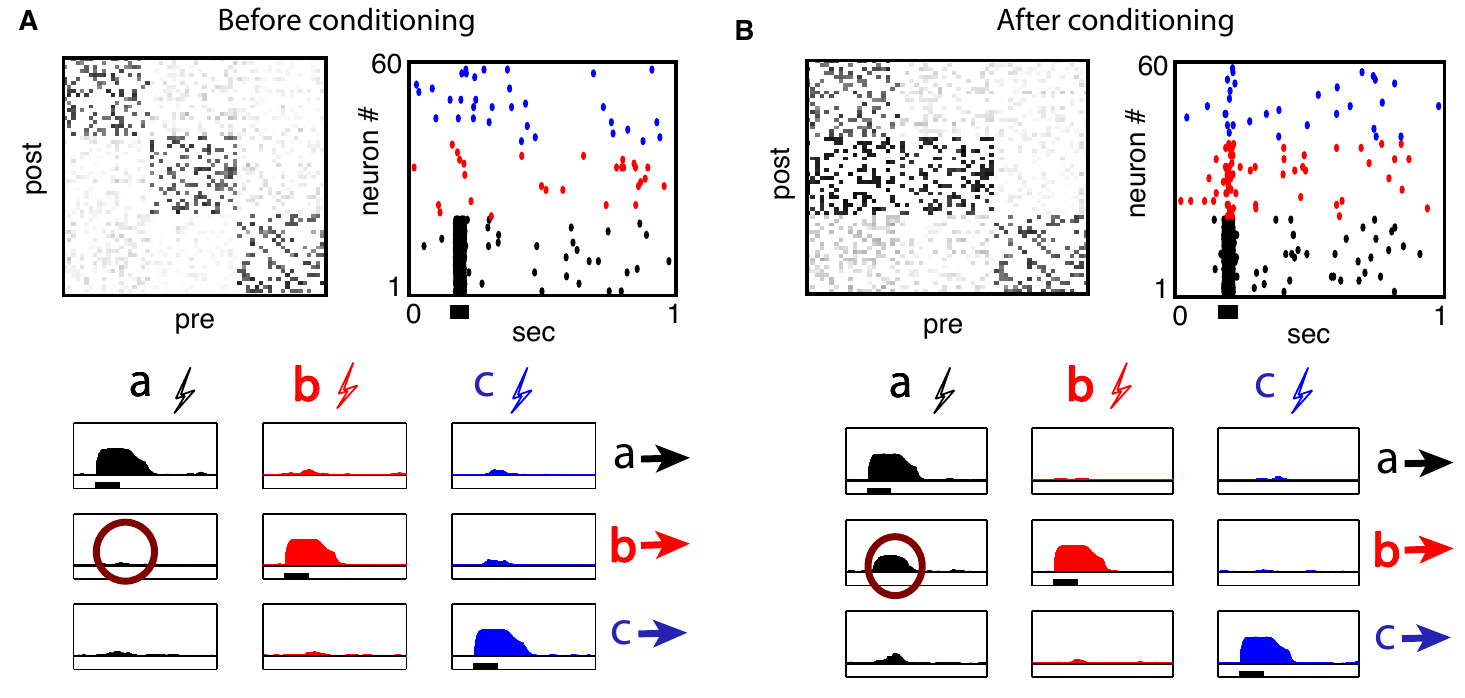}
\caption{({\bf A}) ICMS with baseline connections. ({\bf B}) ICMS after spike-triggered stimulation. Top left panel: Connectivity matrix $J$. Top right panel: Raster plot of entire network as in Figure~\ref{fintro}. Bottom black bar shows a 50 ms stimulation of all neurons in group $a$ at 100 Hz. Bottom: Filtered responses of group projections. Columns designate neural groups being stimulated and rows designate the output evoked from each group that is equivalent to associated EMG responses. }
\label{fig_ICMS}
\end{figure*}
Our model suggests that synaptic changes can indeed occur in MC-like networks, but it remains unclear if such changes can lead to the experimentally observed motor output changes. We address this by simulating EMG responses of our model, before and after spike-triggered conditioning ($\dag$). 

Figure~\ref{fig_ICMS} shows a simulated ICMS protocol before (panel A) and after (panel B) spike-triggered stimulation conditioning. For each case, synaptic matrices are chosen from full network simulations and fixed (STDP is turned off), as shown in the top row of Figure~\ref{fig_ICMS} (reproduced from Figure~\ref{fbci} C). To mimic the ICMS stimulus, we add a square-pulse input rate of 100 Hz lasting 50 ms to the external rate $\nu_\a(t)$ of a target group. An example of the spiking output of our network for $\a=a$ is shown in the top row of Figure~\ref{fig_ICMS} where the solid black bar below the graph shows the stimulus duration. Next, we filter the spike output of all neurons within a group using the synaptic filter $\e(t)$ described earlier, and add them to obtain population activity time-courses. Finally, we take these summed profiles and pass them through a sigmoidal non-linearity ---$(1+\exp{-a(x-b)})^{-1}$ with $x$ the filtered activity--- meant to represent the transform of neural activity to EMG signals. Here, we assume that the hypothetical motoneurons whose EMG is recorded receive inputs only from a single neural group and that network interactions are responsible for cross-group muscle activation.

We choose the nonlinearity parameters $a=2.5$ and $b=5$ to qualitatively reproduce the EMG responses seen in the experiment before spike-triggered conditioning: namely, well-resolved EMG responses are observed only when the relevant MC group is stimulated. The bottom row of Figure~\ref{fig_ICMS} shows simulated EMG responses of each group, averaged over 15 trials each, when ICMS stimulation is delivered to a single group at a time. Panel A shows little cross-group response to ICMS stimulation before spike-triggered stimulation conditioning. However, after conditioning, stimulation of group $a$ evokes an emergent response in the muscle activated from group $b$ (see circles in Figure~\ref{fig_ICMS}), as well as a small increase for group $c$. 

These features were both present in the original experiments (see Figure 2 in~\cite{Jackson:2006kb}) and are consistent with the synaptic strengths across groups before and after conditioning. In addition, the output from ICMS was also measured with isometric torques were produced in directions determined by the evoked EMG responses.  The newly evoked EMG responses after conditioning agree with the observation that torques evoked from $N_{rec}$ (group $a$) typically changed toward those previously evoked from $N_{stim}$ (group $b$). As such, from now on we equate an increase in mean synaptic strength $\bar{J}_{\a\b}$ between groups to an increase in functional connectivity.

\medskip
We conclude that our model satisfies the second experimental observation from~\citep{Jackson:2006kb} we set out to reproduce (point b.~in {\it Introduction}). That is, {\it a simple interpretation of evoked network activity |a filtered output of distinct neural group spiking activity| is consistent with the functional changes in muscle activation in ICMS protocols observed before and after conditioning.}

\subsection{Conditioning effects modulated by realistic activation statistics}

Up to now, we used toy activation profiles $\nu(t)$ in the form of truncated sinusoidal bumps to drive neural activity (Figure~\ref{fintro} B). 
 In this section, we modify our simple model to incorporate experimentally observed cross correlation functions, whenever possible, in an effort to eliminate artificial activation commands and capture more realistic regimes. As a result, we no longer rely on numerical simulations of spiking activity, but rather on analytically derived averaged quantities to explore a wide range of conditioning regimes.
\medskip

Before discussing spiking statistics, we note an important advantage of only considering mean synaptic strengths $\bar J(t)$. For spiking simulations, we use networks of $N=60$ neurons with probability of connection $p=0.3$, which are considerably far from realistic numbers. Nevertheless, the important quantity for mean synaptic dynamics is the averaged summed strengths of synaptic inputs that a neuron receives from any given group: $\frac{pN}{3}\bar J_{\a\b}$. Notice that many choices of $p$ and $N$ can lead to the same quantity, therefore creating a scaling equivalence. Moreover, additional scaling of $J_{max}$ can further accommodate different network sizes.
So far, we assumed that every neuron receives an average of 6 synapses from each group. If each of these synapse were at maximal value $J_{max}=0.1$, then simultaneous spiking from a pre-synaptic group would increase the post-synaptic neuron's spiking probability by $60\%$, a number we consider reasonable.

\subsubsection{Realistic activation statistics}

The changing dynamics of MC neurons in behaving macaques (e.g.,~\cite{Truccolo:2010gz,Carmena:2003bx,Paninski:2004kb,Churchland:2007ht}) yield cross-correlations between cell pairs whose preferred feature can change depending on activity context~\citep{Smith:2009jl,Smith:2009ui,Jackson:2007jq}. 
We do not attempt to capture these subtleties with our simplified model. Rather, we create a family of cross-correlation functions $\hat{C}(u)$ that roughly match averaged quantities observed experimentally, in an attempt to extract qualitative mechanisms. We assume, as was the case so far, that all mean cross-correlations are identical for neurons within the same group and neurons across groups respectively ($\hat{C}_{\a\a}(u)=\hat{C}_{\b\b}(u)$ and $\hat{C}_{\a\b}(u)=\hat{C}_{\gamma\kappa}(u)$).

To calibrate our model, we use estimates of two quantities from MC neuron recordings: the mean cross-correlation between two neurons in the same group and the mean cross-correlation between neurons in different groups. Estimates were obtained from a monkey implanted with a Utah array in primary MC and performing a 2D target tracking task while seated in a chair. See {\it Methods} for details.
 \begin{figure*}[h!]
\includegraphics{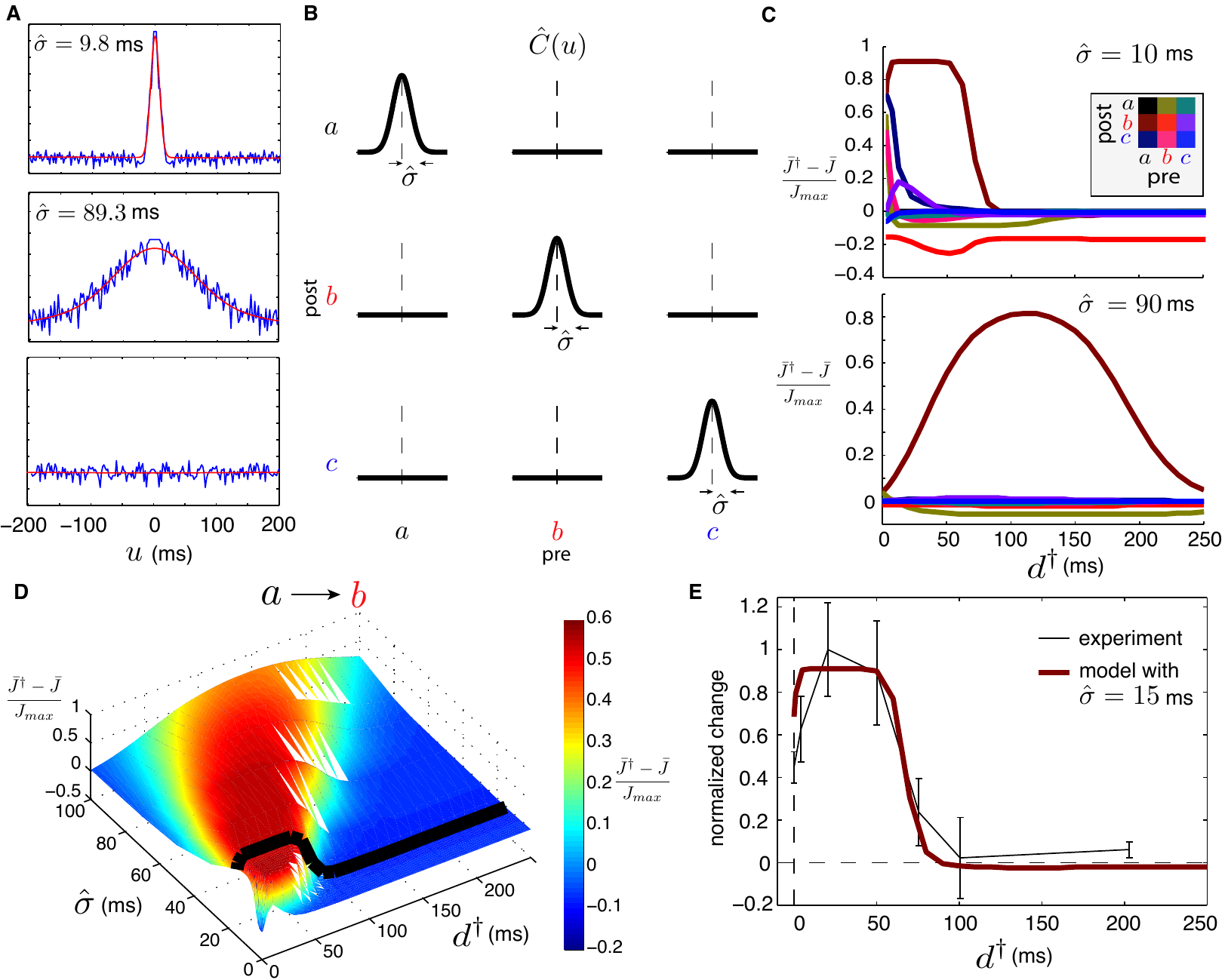}
\caption{({\bf A}) Experimentally obtained cross-correlograms of MC neurons in macaque monkey during a tracking task (blue) and Gaussian fit (red). Top, Middle: for neuron pairs recorded by the same electrode, respectively (using spike-sorting). Top shows the thinnest correlation peak ($\hat{\sigma}=9.8$ ms) and Middle shows the widest ($\hat{\sigma}=89.3$ ms). Bottom: For two neurons recorded by distinct electrodes.
({\bf B}) Cartoon of Gaussian-shaped external cross-correlations $\hat{C}(u)$.
({\bf C}) Relative differences $(\bar{J}^\dag -\bar J)/J_{max}$ for all group-averaged synapses, as a function of stimulation delay $d^\dag$, for the two extremal values of correlation width. Top: $\hat{\sigma}=9.8$ ms. Bottom: $\hat{\sigma}=89.3$ ms.
({\bf D}) Relative differences for averaged synaptic strengths from group $a$ to group $b$ ($(\bar{J}_{ba}^\dag-\bar J)/J_{max}$) as a function of stimulation delay $d^\dag$ and correlation peak width $\hat{\sigma}$. Black line corresponds to best fit plotted in E.
({\bf E}) Superposition of relative difference for $\bar{J}_{ba}$ and normalized mean torque change from spike-triggered conditioning experiments on macaque monkeys. Experimental data from Figure 4 of~\cite{Jackson:2006kb}; error bars show the standard error of the mean. Best fit between model and experimental curve is for $\hat{\sigma}=15$ ms (see black line in D).
}
\label{fpara}
\end{figure*}

In this data set, there were only a few pairs of neurons recorded from the same electrode. Nevertheless, such well-resolved pairs typically showed Gaussian-shaped cross-correlations roughly centered at the origin. An important observation is that the width of the Gaussian peaks varied greatly. Figure~\ref{fpara} A top and middle show both extremes where the cross-correlograms were fitted to Gaussian functions with a standard deviation $\hat{\sigma}$ ranging from $\sim10$ to $\sim90$ milliseconds. In contrast, neuron pairs from different electrodes showed more subtle structure in their cross-correlograms, and a wide variety of feature shapes (e.g.~asymmetrical relationships, anti-correlations, etc.). Most of them showed no correlation features, as depicted by the typical flat cross-correlogram in the bottom of Figure~\ref{fpara} A. An important point in relating these observations to our model is that the mean cross-correlation, across many pairs of neurons within specific groups, is the defining factor that influences synaptic dynamics. While the examples shown in Figure~\ref{fpara} A are insufficient to estimate such averages, they are able to guide parameter ranges in our model.

In light of these observations, we make the following simplifications concerning the model's idealized cross-correlations $\hat{C}(u)$: cross-correlations of neurons within a group are Gaussian while those for neurons across groups remain flat. We get:
\beq
\label{expC}
\hat{C}_{\a\b}(u)=
\begin{cases}
\frac{L}{\hat{\sigma}\sqrt{2\pi}}\exp(-\frac{u^2}{2\hat{\sigma}^2})+B & ; \,\a=\b \\
B & ; \, \a\neq\b
\end{cases}
\eeq
where $B$ is a baseline, $L$ is the height of the peak and $\hat{\sigma}$ modulates the peak's width (i.e. standard deviation). Figure~\ref{fpara} B illustrates our idealized cross-correlation function $\hat{C}(u)$. 
From the analysis in {\it Methods}, it can be shown that the parameters $B$ and $L$, along with $\n$, only influence the speed at which synaptic changes occur (see Equation~\eqref{DS}). Therefore, we set these to obtain a mean firing of about 10 Hz as for the previously assumed activation rates $\nu(t)$, along with $\n$ to obtain convergence timescales similar to those shown in Figure~\ref{fbci} B. We henceforth set $B=0.17$, $L=2.13$ and $\n=10^{-8}$. Since experiments show that $\hat\sigma$ can take a range of values, we leave it as a free parameter in our regime exploration below.

It should be noted that experimentally measured cross-correlations represent the activity of the recorded neurons themselves, as opposed to the cross-correlations of external commands ($\hat{C}(u)$) as we assume here. Nevertheless, we make the simplifying assumption that the resulting timescales of self-consistent network cross-correlations are similar enough to those of external drives to justify the form of $\hat{C}(u)$ used here. Further comments on this assumption can be found in the {\it Discussion}.

\subsubsection{Cross-correlation width influences optimal stimulation delay}
We compute the relative change in mean synaptic strengths ($(\bar J^\dag - \bar J) / J_{max}$) before and after spike-triggered stimulation, for a range of cross correlation widths $\hat \s$ and stimulation delays $\d^\dag$. Figure~\ref{fpara} C shows these differences, for all pre- and post-synaptic neural group combinations, for $\hat \s=10$ ms and $\hat \s=90$ ms.
Notice that the correlation width has significant effects on the $d^\dag$-dependence of BBCI conditioning for all synapses.
For now, we concentrate on the synapses directly targeted by the stimulation procedure, namely those from group $a$ to group $b$ ($\bar J_{ba}$), which naturally show the greatest changes. We discuss the changes incurred by other synapses below.

For both values of $\hat \s$ used in Figure~\ref{fpara} C, the changes incurred by $\bar J_{ba}$ depend non-monotonically on the triggering delay $d^\dag$, admitting either a maximum or a plateau. However the maximizing $d^\dag$ value (or range) changes with $\hat \s$, and so does the sensitivity to choice of $d^\dag$, i.e., how rapidly the curve reaches its maximum as $d^\dag$ is varied. Figure~\ref{fpara} D comprehensively illustrates this dependence, showing $(\bar J_{ba}^\dag - \bar J_{ba}) / J_{max}$ for a range of both $\hat \s$ and $d^\dag$. The main features of this relationship are: (i) optimal delays ($d^\dag$ at peak) get larger as $\hat \s$ grows. (ii) sensitivity is mitigated by larger $\hat \s$, i.e., large correlation timescales lead to optimal delays whose conditioning outcomes are less sensitive to perturbations.

The relative simplicity of our model's equations enables a mechanistic interpretation of these findings. Indeed, Equation~\eqref{function} illustrates that it is the integral of the network's correlations multiplied by the STDP rule that dictates synaptic dynamics. When artificial stimulation is delivered, components of this integral are shifted by $d^\dag$. Wider correlations imply that small changes in shifts $d^\dag$ have slower effects on the integral, leading to both point (i) and (ii) listed above.

When the stimulation of group $b$ is triggered on spikes from the recorded neuron in group $a$, the plasticity incurred by synapses from other neurons in group $a$ to any neuron in group $b$ will depend on how likely it was that $a$-neurons fired within a short interval of the recorded neuron's spikes.
Correlation width measures the synchrony of spiking activity. For narrow correlations, neurons in group $a$ fire closely together and the delay required to induce the maximal plastic change directly depends on the combination of axonal delay and the off-width of the potentiation peak of the STDP rule (see Figure~\ref{fintro2} A in {\it Methods}). Essentially, the peak of the cross-correlation needs to be shifted so it aligns with the peak STDP potentiation.
For wider correlation functions, synaptic changes depend less on synchrony and more on interactions with larger time lags. Indeed, optimal delays are those that shift the cross-correlation's peak past the STDP peak, so that the the latter is aligned with the right-hand-side tails of cross-correlations. This relationship leads to more robust potentiation at larger stimulation delays.

\medskip
This phenomenon constitutes our first novel finding (point 1.~from {\it Introduction}): {\it neuronal activity statistics have an important impact on the value of optimal spike-triggered stimulation delays ($d^\dag$), and can lead to synaptic potentiation that is more robust deviations from that optimal value.}

\subsection{Qualitative reproduction of stimulation/plasticity dependence}
It remains unclear if the mechanisms described above are consistent with the experimentally observed relationship between stimulation delay ($d^\dag$) and efficacy of spike-triggered conditioning in macaque MC. We investigate this by comparing efficacy, as measured by the percentage of torque direction change evoked by ICMS before and after conditioning~\citep{Jackson:2006kb}, to relative synaptic strength changes in our model. This is motivated by the above demonstration that synaptic strengths are well correlated with amplitude of evoked muscle activations in a ICMS experiment (see Figure~\ref{fig_ICMS} and point b. in {\it Introduction}).
Nevertheless, the following comparison between model and experiment is qualitative, and meant to establish a correspondence of (delay) timescales only.

We use the data originally presented in Figure 4 of~\cite{Jackson:2006kb}, describing the shift in mean change in evoked wrist torque direction by ICMS of the $N_{rec}$ site (analogous to our group $a$), as a function of stimulation delay $d^\dag$. We plot the same data in Figure~\ref{fpara} E, with the maximal change (in degrees) normalized to one. On the same graph, we plot the $(\bar J_{ba}^\dag - \bar J_{ba} )/ J_{max}$ v.s. $d^\dag$ curve for the value of $\hat \s$ that offers the best fit (in L1-norm). This amounts to finding the best ``$\hat \s$-slice" of the graph in Figure~\ref{fpara} D to fit the experimental data. We found that $\hat \s \simeq 15$ ms gives the best correspondence.

We reiterate that this comparison is qualitative. Nevertheless, the fit between the $d^\dag$-dependence of experimentally observed functional changes and modelled synaptic ones is clear.  As our model's spiking activity and STDP rule are calibrated with experimentally observed parameters (see {\it Methods}), this evidence suggests that our simplified framework is consistent with experiments.
Importantly, $\hat \s=15$ ms is comparable to correlation timescales between functionally related MC neurons in macaque during a stereotyped motor task~\citep{Smith:2009jl,Smith:2009ui,Jackson:2003tx}, where such correlations are reported to be gaussian-like and have peak width at half height on the order of 20 ms, corresponding roughly to $\hat \s=10$ ms. It is generally believed that task-specific motion introduces sharp correlations and that free behaving, rest and sleep states induce longer-range statistics~\citep{Jackson:2007jq}. The conditioning experiment in~\cite{Jackson:2006kb} was conducted over a wide range of states, including sleep, which may lead to longer mean cross-correlation timescales. 
A prediction of our model is that spike-triggered conditioning restricted to periods of long timescale correlations in MC, such as during sleep~\citep{Jackson:2007jq}, could lead to a more robust conditioning dependence on stimulation delays (see {\it Discussion}).

\medskip
This finding implies that our model successfully reproduces the third and last experimental observation from~\cite{Jackson:2006kb} (point c.~from {\it Introduction}): {\it using simplified cross-correlation functions of MC neural populations calibrated from experimental measurements, our model reproduces the relationship between the magnitude of plastic changes and the stimulation delay in a spike-triggered conditioning protocol.}

\subsection{Multi-synaptic and collateral effects.}
\begin{figure*}[h!]
\includegraphics{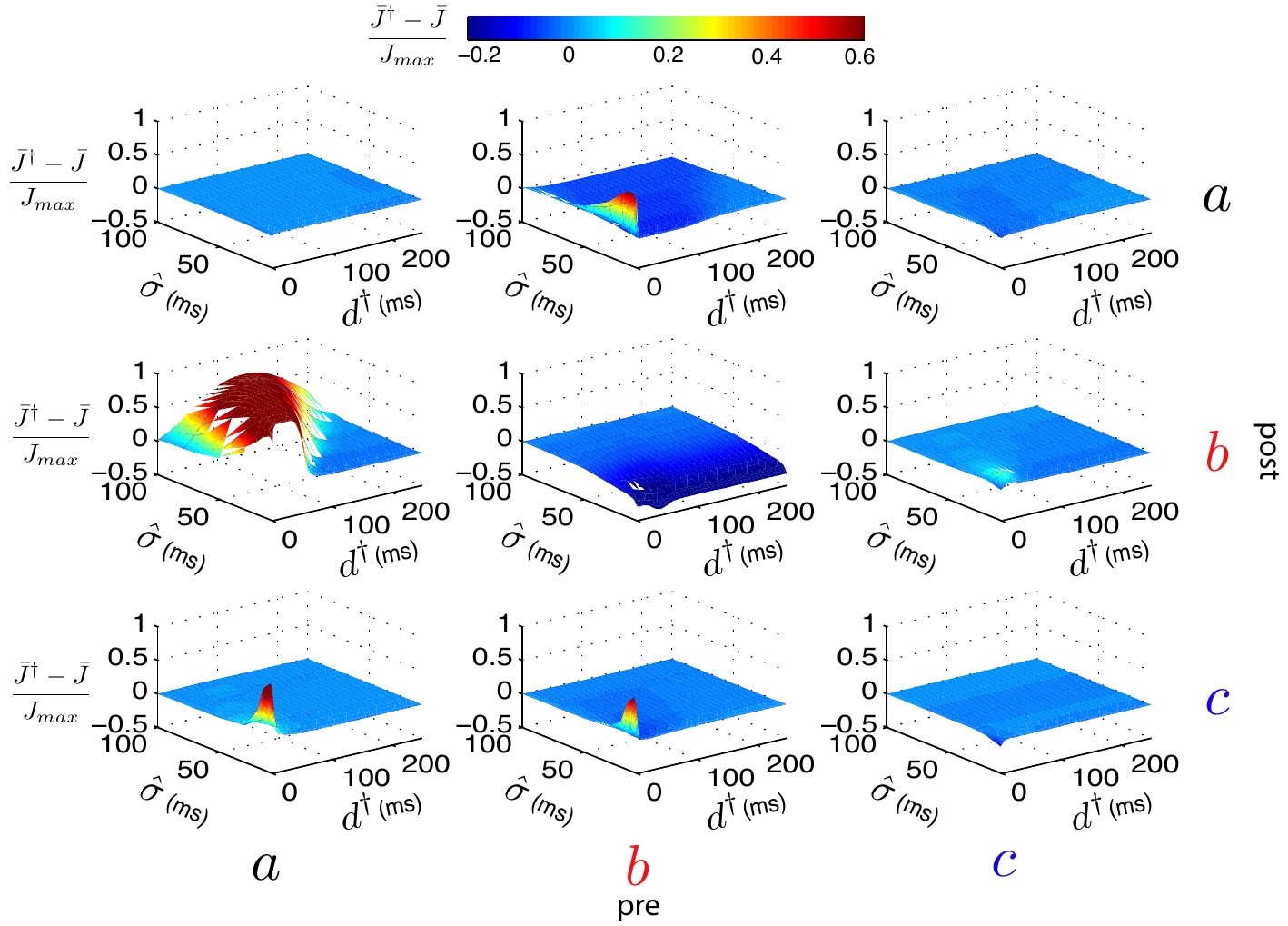}
\caption{
Relative differences for mean synaptic strengths $(\bar{J}_{\a\b}^\dag-J)/J_{max}$ as a function of stimulation delay $d^\dag$ and correlation peak width $\hat{\sigma}$, for all combinations of pre- and post-synaptic groups $a$, $b$, $c$.
}
\label{fgrid}
\end{figure*}

We now explore the effects of spike-triggered stimulation on collateral synaptic strengths, i.e., other than the targeted $a$-to-$b$ pathway ($N_{rec}$-to-$N_{stim}$). Figure~\ref{fgrid} shows color plots of these changes as a function of $d^\dag$ and $\hat \s$, for the nine combinations of pre- and post-synaptic groups.

For a wide range of parameters, there is little change other than for the $a$-to-$b$ synapses. Indeed, when external cross-correlation width $\hat \s$ is moderate to large, spike-triggered stimulation has little effect on collateral connections, for any stimulation delay. This is in conjunction with the robustness of $a$-to-$b$ changes discussed in the previous section.
 Nevertheless, some localized features arise when the external cross-correlation width $\hat \s$ is small.

\medskip
First, $b$-to-$b$ synapses become depressed, regardless of stimulation delay, for short correlation timescales. This is due to the occurrence of synchronized population spiking produced by artificial stimulation which, because of our choice of dendritic and axonal delays (see {\it Methods}), promote depression. Such synchronized spikes induce sharp $\d$-peaks in the network's cross-correlation (see Figure~\ref{fcorrs} B in {\it Methods}) and the combination of transmission delays shifts this peak toward the depression side of the STDP rule. When external cross-correlations are narrow (i.e. small $\hat \s$), their interaction with the STDP rule --which manifests in the integral in Equation~\eqref{mean:DJ}-- is more sensitive to the addition of such $\d$-peaks, resulting in overall depression. In contrast, when cross-correlations are wider, the addition of $\d$-peaks has a smaller effect since a wider range of correlations contribute to the integrated STDP changes.

\medskip
Second, the synapses from $b$ to $a$,  become potentiated for short delays $d^\dag$ when $\hat \s$ is small enough. 
This happens because of a combination of factors. When the recorded neuron in group $a$ spikes, the population-wide spike artificially elicited in $b$ quickly follows and travels to the $b$-to-$a$ synapses. This means that the spike of a single neuron in $a$ affects all neurons in $a$, with an effect amplified by the strength of many synapses, shortly after the neuron originally fired. When cross-correlations among $a$-neurons are wide, the effect of this mechanism is diluted, similarly to the $b$-to-$b$ synapses discussed above. However, when $a$ is nearly synchronous, this short-latency feedback produces synaptic potentiation of the $b$-to-$a$ synapses.

\medskip
Lastly, the synapses from both groups $a$ and $b$ onto the control group $c$ are also potentiated when both $\hat \s$ and $d^\dag$ are small enough. This can be explained in two parts and involves di- and tri-synaptic mechanisms. When the recorded neuron in $a$ fires a spike, a population-wide spike is artificially evoked in $b$ shortly after, which travels down to $b$-to-$c$ synapses and elicits a response from neurons in $c$. Narrow cross-correlations imply that many spikes in $a$ fall within a favorable potentiation window of spikes in $c$, thereby contributing to the potentiation of $a$-to-$c$ synapses. 
In turn, the strengthening of $a$-to-$c$ synapses imply that spikes in $a$ are more likely to directly elicit spikes in $c$, thereby repeating the same process in a different order for $b$-to-$c$ synapses. Note that without $a$-to-$c$ synapses, the $b$-to-$c$ synapses would not potentiate. Indeed, as was the case for $b$-to-$b$ synapses, the combination of transmission delays do not conspire to promote direct potentiation following a population-wide synchronous spike. 

This illustrates that emergent, multi-synaptic mechanisms give rise to significant changes in collateral synapses, and that these mostly occur for short stimulation delays and short correlation timescales.

\medskip
This constitutes our second novel finding (point 2.~from the {\it Introduction}): {\it multi-synaptic mechanisms give rise to significant changes in collateral synapses during conditioning with short delays, and these cannot be attributed to STDP mechanisms directly targeted by the BBCI, instead emerging from network activity}.

\subsection{Stimulating a subgroup of neurons}

We have assumed that spike-triggered stimulation elicits population-wide synchronous spiking of all neurons in group $b$ ($N_{\rm stim}$ in~\cite{Jackson:2006kb}). 
This is valid if the neural group $b$ represents all the neurons activated by the stimulating electrode, but is not necessarily representative of the larger population of neurons that share external activation statistics due to a common input $\nu_b(t)$. 
Indeed, some neurons that activate in conjunction with those close to the electrode may be far enough from it so they do not necessarily spike in direct response to a stimulating pulse.
Alternatively, selective activation of neurons within a group can also be achieved via optogenetic stimulation in a much more targeted fashion~\citep{Ledochowitsch:2015ga}.
We now consider the situation in which only a certain proportion of neurons from group $b$ is activated by the spike-triggered stimulus. 

\begin{figure*}[h!]
\includegraphics{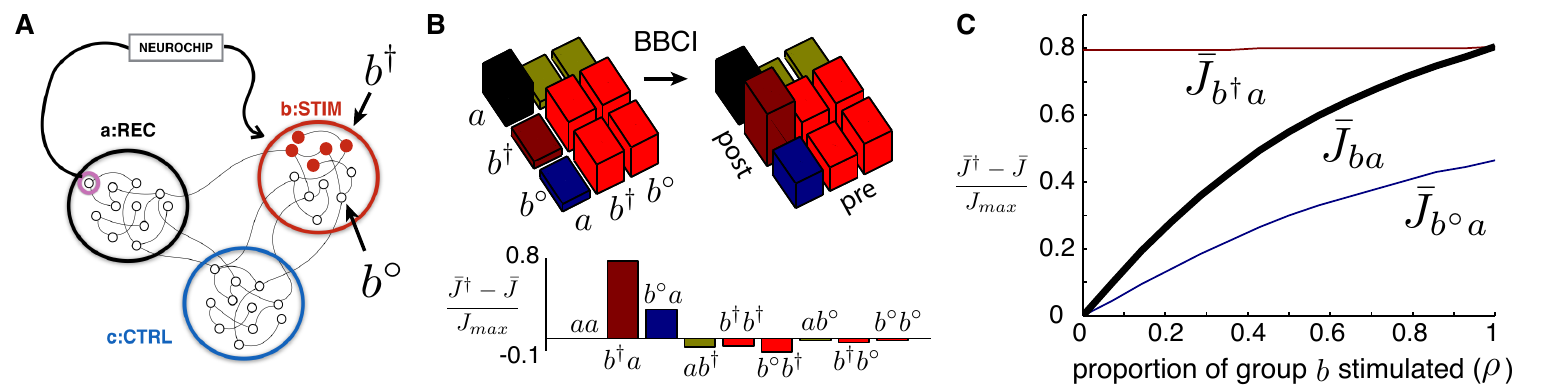}
\caption{({\bf A}) Network with group $b$ subdivided into two subgroups: stimulated neurons ($b^\dag$, red) and unstimulated neurons ($b^\circ$, white).
({\bf B}) Top: plots of equilibrium synaptic strengths $\bar{J}^*$ between group $a$ and both subgroups of $b$ in normal activity and under spike-triggered stimulation (left/right resp.). Proportion of stimulated neurons is set to $\r=0.5$. Bottom: Plot of relative synaptic changes between equilibria: $(\bar{J}_{\a\b}^\dag-\bar{J}_{\a\b})/J_{max}$. 
({\bf C}) Relative synaptic changes as a function of proportion of neurons in group $b$ receiving stimulation $\r$. Plotted are subgroup averaged synaptic changes for synapses from $a$ to $b^\dag$ and $b^\circ$ and from $a$ to entire group $b$.
}
\label{fprop}
\end{figure*}

\medskip
We denote the stimulated subgroup by $b^\dag$ and the un-stimulated subgroup by $b^\circ$ (Figure~\ref{fprop} A). 
All neurons in group $b$ receive the same external rates $\nu_b(t)$ as before, but only a few (solid red dots) are stimulated by the Neurochip.
Let $N_b=N/3$ be the number of neurons in group $b$ and the parameter $\r$, with $0\leq \r\leq 1$, represent the proportion of stimulated neurons in $b$. The sizes of groups $b^\dag$ and $b^\circ$ are given by $N_b^\dag=\r N_b$ and $N_b^\circ=(1-\r)N_b$, respectively. 
We now adapt our analytical averaged model~\eqref{function} to explore the effect of stimulation on subdivided synaptic equilibria. We verified that the analytical derivations used below match the full spiking network simulations as before (data not shown).

Both subgroups of $b$ receive the external rate $\nu_b(t)$ but only one receives spike-triggered stimulation. These changes are captured in the averaged analytical derivations by tracking the number of neurons in each sub-group in the averaging steps leading to Equations~\eqref{avr_corrections} and~\eqref{function} accordingly --- replacing $N/3$ by $N^\dag_b$ and $N^\circ_b$ where necessary. This way, we obtain subgroup-specific synaptic averages (e.g.~$\bar J_{b^\dag a}$).

Figure~\ref{fprop} B shows the group-averaged connectivity strength between group $a$ (the recording site) and both subgroups of $b$, before and after spike-triggered stimulation. External cross-correlations $\hat C(u)$ are as in Figure~\ref{fpara} B with $\hat \s=20$ms, and stimulation delay is set at $d^\dag=30$ms. The proportion of stimulated neurons in $b$ is set to $\r=0.5$. The bottom of the same panel shows the normalized changes of mean synaptic strengths due to spike-triggered stimulation. 
As established for the original network (see Figure~\ref{fbci} B), the biggest change occurs for synapses from $a$ to the subgroup that is directly stimulated ($b^\dag$). However for subgroup $b^\circ$, we see a noticeable change in its incoming synapses from group $a$, in contrast to synapses of other un-stimulated groups ($c$) that do not appreciably change (not shown). This means that sharing activation statistics with stimulated neurons is enough to transfer the plasticity-inducing effect of conditioning to a secondary neural population.

Next, we investigate how this phenomenon is affected by the proportion of neurons in $b$ that receive stimulation: $\r$. Figure~\ref{fprop} C shows the subgroup-averaged normalized changes of synapses from group $a$ to subgroups $b^\dag$ and $b^\circ$, as $\r$ varies between 0 and 1. When more neurons get stimulated, the transferred effect on the unstimulated group is amplified. This means that the combined outcome on the entirety of group $b$ grows even faster with $\r$ (supra-linearly), as shown in Figure~\ref{fprop} C, where the combined $b$-averaged changes in synaptic strength, $\bar J_{ba}=\r \bar J_{b^\dag a}+(1-\r) \bar J_{b^\circ a}$, are plotted as a function of $\r$. 

\medskip
In summary, this phenomenon represents our third and final finding (point 3.~in {\it Introduction}). Our model shows that
{\it neurons not directly stimulated during spike-triggered conditioning can be entrained into artificially induced plasticity changes by a subgroup of stimulated cells, and that the combined population-averaged effect grows supra-linearly with the size of the stimulated subgroup.}

\section{Discussion}

\subsection{Summary}

In this study, we used a probabilistic model of spiking neurons with plastic synapses obeying a simple STDP rule to investigate the effect of a BBCI on the connectivity of recurrent cortical-like networks. 
Here the BBCI records from a single neuron within a population and delivers spike-triggered stimuli to a different population after a set delay.
 We developed a reduced dynamical system for the average synaptic strengths between neural populations; these dynamics admit stable fixed points corresponding to synaptic equilibria that depend solely on the network activity's correlations and spike-triggered stimulation parameters. In this framework, individual synapses may fluctuate with ongoing network activity but their population average remains stable in time. We validate our findings with detailed numerical simulations of a spiking network and calibrate our result based on experiments in macaque MC. To our knowledge, this is the first time a plastic spiking network model includes recurrent network interactions to capture the effects of a BBCI.

We successfully reproduce key experimental results from~\citep{Jackson:2006kb} that describe synaptic changes due to spike-triggered conditioning. 
Specifically, we recover the two emergent timescales with which these changes occur (hours to days), we show that filtered evoked activity  from our network mimics muscle EMG evoked by ICMS protocols, and we show that maximal changes in mean synaptic strength from the recording site to the stimulated site occur with stimulation delays in the 20-50 ms range, as was the case for experiments.
Furthermore, we formulate three novel findings using our model. We outline the relationship between temporal statistics within neural populations and optimal stimulation parameters, we uncover multi-synaptic mechanisms that emerge from spike-triggered conditioning that are not directly predicted by a single-synapse STDP rationale, and we found that the stimulation of a subset of neurons within a population can lead to supra-linear scaling of effects.

Based on this, we formulate two main experimental predictions:
\begin{enumerate}
\item Spiking statistics in a network greatly influence the effects of stimulation, with a wider range of spike-stimulation delays leading to optimal evoked plasticity when correlations have timescales. We predict that restricting spike-triggered stimulation to states with these statistics, such as sleep, or potentially driving such correlation artificially, may provide more robust outcomes.
\item For very short delays, stimulation may produce collateral effects in other synaptic connections due to polysynaptic interactions. We propose that under special conditions, it may be possible to influence collateral synaptic pathways that are not directly targeted by a spike-triggered stimulation protocol.
\end{enumerate}

\subsection{Cortical activity: driven vs emergent}

In our model, input rates are needed to endow neural groups with desired statistics.
In cortical networks it is unclear how much of observed activity is driven from external sources and how much is due to intra-network interactions. 
We argue that the use of prescribed external activations is appropriate since we show that plasticity works to ``learn" activation correlations (see also~\cite{Gilson:2010eh,Gilson:2009iy,Gilson:2009gu,Gilson:2009eq,Gilson:2009ci}), thereby guiding synaptic connections to promote the same spontaneous network statistics as its driven ones.
Therefore, while external input rates are necessary in our model, the resulting network correlations $C(u)$ reflect both emergent and external dynamics, and serve as a proxy for MC activity.

\subsection{State-dependent stimulation: designing adaptive stimulation strategies}

An interesting outcome of our findings is that spike-triggered conditioning is strongly influenced by the source and target cross- and auto-correlation structure $C(u)$. Throughout the paper, we assumed that these statistics were stationary for the entire simulated period. However, it is well known that cortical circuits can show a wide range of activity regimes depending on the state of the animal.
The statistics of MC neurons may be very different if the animal is awake and behaving freely or performing a precise motor task, or is asleep~\citep{Jackson:2007jq}.
While such states have limited durations, their timescales (from minutes to hours) may be long enough to define locally-stationary statistics that BBCI protocols could leverage to optimize desired effects. For example, stimulation during sleep, which is known to produce oscillation-rich activity with longer-range correlations~\citep{Bazhenov:2002fn,Jackson:2007jq}, could have more robust but slower effects, while stimulation during a specific task can have more targeted outcomes.

\subsection{Scalable framework: from experimental to clinical applications}

Our model can easily be scaled up to include multiple recording and stimulation sites, and different stimulation protocols, including, e.g. EMG-triggering~\citep{Lucas:2013wj} or paired-pulse stimulation~\citep{Seeman:2015wn}. 
It is also easily adaptable to optically-based stimulation which can target specific neurons within functional groups (see e.g.~\cite{Ledochowitsch:2015ga,Silversmith:tu,CarrilloReid:2016jn}).
As it does not require costly simulations |only theoretical estimates| it is straightforward to apply optimization algorithms to find the best stimulation protocol to achieve a desired connectivity between cortical sites.
Furthermore, it can easily incorporate closed-loop signals such as changes in recorded statistics in real-time.
This framework offers a flexible testbed to help design experiments and clinical treatments.

Neural implants such as BCIs and BBCIs are under active development as they have significant potential for clinical use~\citep{Potter:2014ig}.
 Among other outcomes, they are promising avenues for treatment of motor disabilities.
Indeed, a BBCI capable of inducing plastic changes in cortical circuits could be used to promote novel synaptic pathways in order to restore functional connectivity after a stroke or injury~\citep{Edwardson:2013uc,Guggenmos:2013fq}.

For such bidirectional neural implants to be successful, a number of real-time computational issues need to be resolved. 
Our modeling framework presents an important step toward development of rapidly computable guides for controllers, based on anatomical organization, measurable network properties and known physiological mechanisms such as STDP.

\subsection{Model limitations and future steps}

While being able to capture important features, our model misses some important physiological aspects of MC circuits.
Nevertheless, we note that the simplicity of our model is
a strength: it captures complex network-level plasticity changes with only excitatory activity and STDP. This suggests that excitatory mechanisms are central to artificially-induced plasticity by a BBCI. 
 However, this simplicity will likely be insufficient to reproduce more complex protocols that also recruit inhibitory populations. An example is paired-pulse conditioning, where the BBCI stimulates several neural sites, with different time delays. Additional biological realism is key to expand our bottom-up theoretical framework. 

A number of steps can be taken to add physiological realism, each of them adding some complications. 
For inhibition, there are technical issues when considering probabilistic spiking and inhibition (e.g. inhibition can induce ``negative" spiking probabilities if unchecked), and STDP mechanisms for inhibitory synapses are not well understood.
The implementation of our framework in a dynamical model setting, building on theoretical models of inhibitory plasticity (see e.g.~\cite{Vogels:2011er}) are natural next steps. 
In addition, the inclusion of spatial structure and heterogeneous delay distributions, based on anatomical data~\citep{Markram:2015dq}, is necessary to expand the framework to multiple cortical sites and spinal cord. 
Finally, the inclusion of modulatory mechanisms, activity- or chemically-dependent, is crucial to capture phenomena such as synaptic consolidation and adaptation. 

\medskip
In summary, our reduced dynamical system approach is a promising basis upon which to build and ultimately to predict the effects of finer-grained cell-type specific and temporally structured activation patterns afforded by next-generation neural implants using electrical or optical stimulation.

\section{Methods}

This section contains details about our model, numerical simulations and analytical derivations of averaged synaptic dynamics. 
We closely follow prior literature~\cite{Gilson:2010eh}, and develop key innovations to better suit our model's BBCI components.

\subsection{Model}

\subsubsection{Spiking activity}
We consider a network of $N$ model neurons whose underlying connections are
given by an $N\times N$ connectivity matrix $J(t)$ with synaptic weights that evolve according to a plasticity rule (as described below).
The presence of a directed synapse $J_{ij}$ from neuron $j$ to neuron $i$ is randomly and independently determined with probability $p$.
Throughout this paper, we consider sparse networks with $p=0.3$.
Once an initial connectivity matrix $J(0)$ is drawn, existing synapses are allowed to change but no new synapses can be created. All synapses are excitatory so that $J_{ij}\geq0$.

The spike train of neuron $i$ is a collection of times $\{t^s_i\}$ and can be written as sum of Dirac-$\d$ functions: $S_i(t)=\sum_{t^s_i}\d(t-t^s_i)$. We are interested in the probability of observing a certain spike train $S_i(t)$ given a certain connectivity matrix $J(t)$, the
other neurons' activity and the external driving signal $\nu_i(t)$.
We assume that this spiking probability can be expressed as a time-dependent density $\l_i(t)$, the parameters of which are governed by the {\it ensemble average} $\la S_i(t) \ra$ and must be consistent across the network.
For simulations, we assume $\l_i(t)$ is a Poisson rate, but most of the derivations we make below are generalizable to other probabilistic or dynamical models of network spiking activity. For example, an {\it integrate-and-fire} spiking mechanism could be used, with a stochastic transfer function derived from Fokker-Planck equations using linear response theory, as is done in~\citep{Lindner:2005wx,Ocker:2015do}.

The instantaneous firing rate at time $t$ for the $i^\text{th}$ neuron in the network, subject to an external driving rate $\nu_i(t)$, is given by
\beq
\label{model}
\begin{split}
 \l_i(t)&=\nu_i(t)+\sum_j \sum_{t_j^s} J_{ij}(t) \e(t-t_j^s-d^a_{ij}).
\end{split}
\eeq
where $$\e(t)=H(t)\frac{1}{\tau}e^{-t/\tau}$$ is a synaptic filter with $H$ denoting the Heaviside function and $\tau$ a synaptic time constant which we set to $\tau=5$ms. This filter is normalized  ($\int_{-\infty}^{\infty}\e(t)dt=1$) and causal ($\e(t)=0$ for $t<0$) so that only past spikes influence spiking probabilities at $t$. Finally, we introduce axonal delays, independently sampled for each synapse from a uniform distribution $d^a_{ij}\sim\bar{d}^a\pm\sigma^a$ with $\bar{d}^a= 3$ ms and $\sigma^a=1$, a range consistent with experimental studies (see e.g.~\cite{Markram:1997hd,Markram:2015dq}).
We simulate~\eqref{model} numerically by discretizing time in small bins of size $\d t$ and independently drawing spikes for neuron $i$ in bin $[t,t+\d t]$ according to the probability $\l_i(t)\d t$ (see {\it Simulation details} below).

\begin{figure*}
\includegraphics{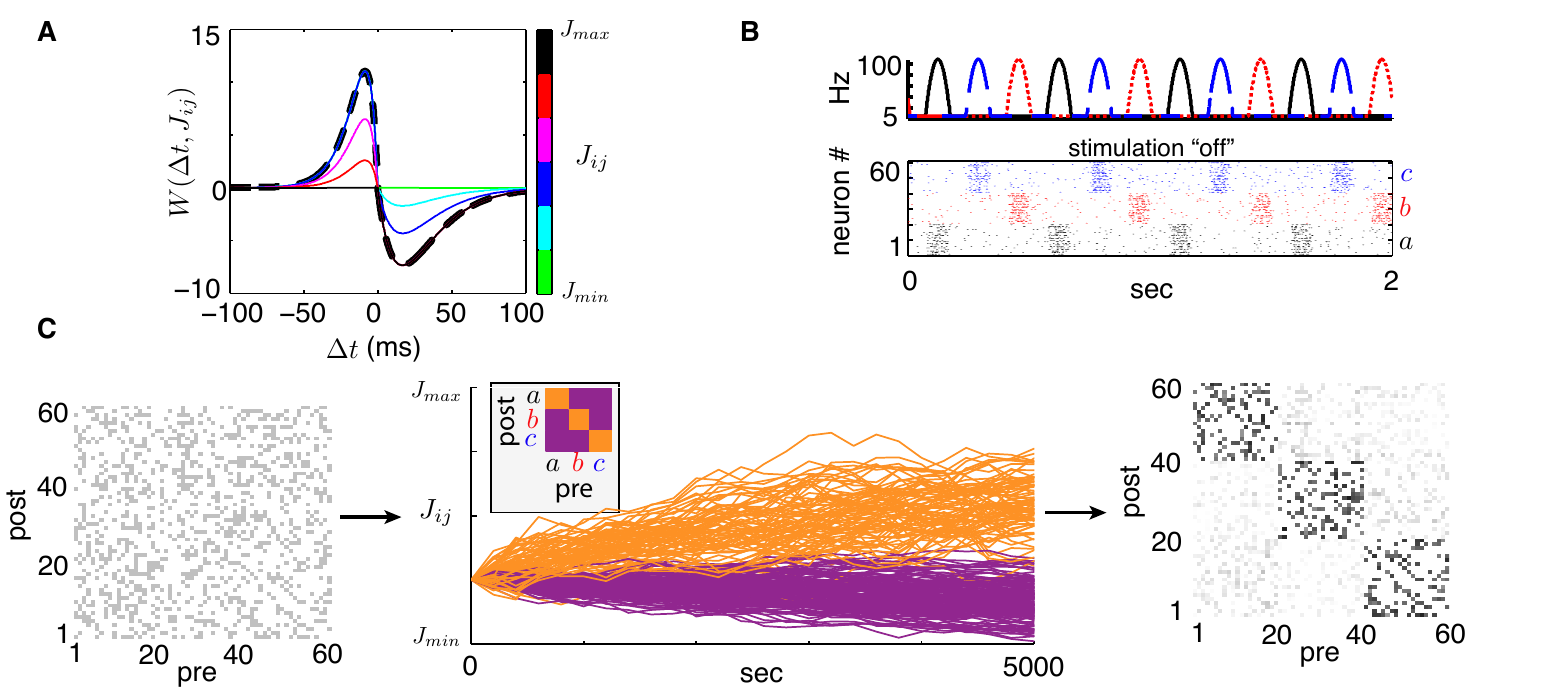}
\caption{
({\bf A}) STDP function $W(\D t,J_{ij})$. Dashed black line indicates the $W^+$ and $W^-$ functions. Coloured lines indicate the full, weight-dependent rule $W$ for different synaptic weight values $J_{ij}$ shown in the color bar.
({\bf B}) Top: Periodic external rates delivered to each neural group (as in Figure~\ref{fintro}). Middle: sample spike raster plot from a simulation subject to rates plotted above with sparse, homogeneous connectivity $J(0)$ ($N=60$).
({\bf C}) Simulated synaptic evolution $J(t)$ with $J_{min}=0$, $J_{max}=0.1$ and $\n=10^{-8}$ under baseline conditions (stimulation ``off") as in B. Left: initial connectivity matrix with connectivity probability $p=0.3$ and $J_{ij}(0)=J_{max}/4$ (grayscale: white $=J_{min}$, black $=J_{max}$). Middle: 10 evolving weights $J(t)$ randomly selected from each of 9 possible pre- and post-synaptic group combinations. Color indicates whether a synapse connects neurons within the same group (orange) or across different groups (purple). Right: synaptic matrix $J(t)$ at the end of the simulation, once group-averaged weights have reached steady-states.}
\label{fintro2}
\end{figure*}

\medskip
\noindent \textbf{\textit{Plastic synapses.}}
For each pair of pre- and post-synaptic spike times $t_{pre}$ (from neuron $j$) and $t_{post}$ (from neuron $i$), the synapse $J_{ij}$, if present, will be changed according to a STDP rule $W$ defined below. This rule defines increments that are added to existing synaptic weights, which are updated every time a new spike is fired. We describe this update procedure below (see Equation~\eqref{model:plastic}). For an inter-spike-interval $\D t=t_{pre}-t_{post}$ at a synapse with weight $J_{ij}$, this increment is given by:
\beq
\label{STDP}
W(\D t, J_{ij})=
\begin{cases}
   f^+(J_{ij})W^+(\D t) & \D t<0 \\
  -f^-(J_{ij})W^-(\D t) & \D t > 0
\end{cases}
\eeq
where $\D t=t_{pre}-t_{post}$. This is a {\it multiplicative} STDP rule since it is a product of a weight-dependent term $f^\pm(J_{ij})$ and a spike-time-dependent term $W^\pm(\D t)$. Figure~\ref{fintro2} A shows a plot of $W(\D t,J_{ij})$ as a function of $\D t$ for various values of $J_{ij}$.

The functions $W^+(\D t)$ and $W^-(\D t)$ describe the potentiation and depression of a synapse, respectively, as a function of spike-timing. We assume Hebbian plasticity and require that these functions be both positive and decay to zero as $|\D t|$ grows, so that a pre-synaptic spike preceding a post-synaptic one leads to potentiation while the opposite leads to depression, in accordance with experimental estimates for excitatory STDP~\citep{Bi:2001uy,Caporale:2008fm,Markram:1997hd}. Many functions $W^\pm$ may lead to similar results; here we choose

\beq
\label{W}
\begin{split}
W^+(\D t)&=A^+\frac{|\D t|}{\tau^+}\exp(-|\D t|/\tau^+)\\
W^-(\D t)&=A^-\frac{|\D t|}{\tau^-}\exp(-|\D t|/\tau^-)
\end{split}
\eeq
as in~\citep{Kempter:1999tn,Gilson:2010eh} and others where $A^+$ and $A^-$ indicate maximal increments and $\tau^+$ and $\tau^-$ are decay time constants. Experimental evidence suggests that $W^+$ and $W^-$ are not identical so that STDP is not symmetric, with the maximum potentiation increment often being larger than its depression counterpart ($A^+>A^-$), and decay lengths following opposite trends ($\tau^->\tau^+$)~\citep{Bi:2001uy}.
As in~\citep{Kempter:1999tn,Gilson:2010eh}, respecting generally accepted ranges from various animal models~\citep{Wang:2005fj,Bi:2001uy,Bi:1998ve,Caporale:2008fm},
we set: $A^+= 30$, $A^-= 20$, $\tau^+= 8.5$ ms, $\tau^-=17$ ms.

The terms $f^+(J_{ij})$ and $f^-(J_{ij})$ control the magnitude of plastic increments as a function of synaptic strength and decay to zero as $J_{ij}$ approaches set bounds $J_{max}$ and $J_{min}$ respectively.
Thus, synaptic weights can only asymptotically approach these bounds under Equation~\eqref{STDP}.
Here, we choose these functions to be

\beq
\label{J-mult}
\begin{split}
f^+(J_{ij})&=\left(1-\frac{J_{ij}}{J_{max}}\right)^\g\\
f^-(J_{ij})&=\left(\frac{J_{ij}}{J_{max}}\right)^\g
\end{split}
\eeq
where the exponent $\g$ controls the strength of the weight dependence. We set $\g=0.1$, $J_{min}=0$, $J_{max}=0.1$ for our simulations and discuss maximal synaptic strengths in the {\it Theory} section below.
This weight-dependent mechanism reflects the fact that synapses cannot be infinitely strong and that as a synapse potentiates, it becomes harder to potentiate it further as resources to do so deplete (e.g. availability of neurotransmitters and vesicles, etc.).
 This is consistent with experimental findings~\citep{Wang:2005fj,Bi:1998ve}.
Rather than introducing unrealistic rate-dependent plasticity terms and/or imposing periodic rescaling of synapses, as is often done in the modelling literature, the use of weight-dependent terms in Equation~\eqref{STDP} will enable stable connectivity equilibria, akin to homeostasis, while remaining biologically plausible. \bigskip

In addition to axonal delays $d^a_{ij}$ introduced earlier, we add dendritic delays independently sampled from a uniform distribution $d^d_{ij}\sim\bar{d}^d\pm\sigma^d$ with $\bar{d}^d=2$ ms and $\sigma^d=1$ as an estimate of physiological values (see e.g.~\citep{Markram:1997hd}).
These account for the time for a spike in neuron $i$ to travel back up dendrites and trigger STDP effects for its synapses. Therefore, for every pair of spike times ($t_i^s$, $t_j^s$), the synapse $J_{ij}$ will be updated at time $t^*=\max(t^s_i+d^d_{ij},t^s_j+d^a_{ij})$:
\beq
\label{model:plastic}
J_{ij}(t^*)\to J_{ij}(t^*)+\n W([t^s_j+d^a_{ij}]-[t^s_i+d^d_{ij}],J_{ij}(t^*))
\eeq
where $\n\ll1$ scales
the overall STDP learning rate and is set to $\n=10^{-8}$ for simulations. 
It is chosen by directly comparing our model's predicted timescale for synaptic changes and the BBCI experiment we aim to capture~\citep{Jackson:2006kb} (see also {\it Results}).

\subsubsection{Functional groups and spike-triggered stimulation}

As illustrated in Figure~\ref{fintro} A and outlined in {\it Results}, we separate our network in three equally sized groups ($a$, $b$ and $c$) in which neurons receive the same external drive $\nu_\a(t)$ ($\a=a,b$ or $c$). These groups are indexed in order so that neurons $i=1,...,N/3$ are in group $a$, etc. 
For exposition, Figure~\ref{fintro2} C shows time-traces of the synaptic weights $J(t)$, with $J(0)$ randomly initiated as a sparse matrix with homogeneous weights $J_{init}=J_{max}/4$. We can already see some connectivity structure emerge from network activity, with synapses that connect neurons within the same group (in orange) gradually increasing and synapses connecting neurons of different groups (in purple) decreasing.

To mimic the spike-triggered stimulation experiments in~\cite{Jackson:2006kb}, we assign group $a$ to be the ``Recording" site, group $b$ to be the ``Stimulation" site and $c$ to be the ``Control" site. 
Neuron $i=1$ from group $a$ is the ``recorded" neuron, whose spikes trigger stimulation of every neuron in group $b$ after a stimulation delay $d^\dag$.
Stimulation of subsets of b is also considered in {\it Results}.
Thus, we impose the following interactions:
\beq
\label{model_stim}
\begin{cases}
\l^\dag_i(t) = \nu_{b}(t)+\sum_jJ_{ij}(t)\sum_{t^s_j}\e(t-t_j^s-d^a_{ij})+\sum_{t^s_1}\d(t-t^s_1-d^\dag) &; \quad i\in b\\
\l^\dag_i(t) = \nu_{\a_i}(t)+\sum_jJ_{ij}(t)\sum_{t^s_j}\e(t-t_j^s-d^a_{ij})&; \quad i\notin b
\end{cases}
\eeq
where ``$\dag$" indicates the presence of stimulation.
The bottom panel of Figure~\ref{fintro} B shows the simulated spiking activity of a network with this stimulation protocol turned ``on" and a delay $d^\dag=10$ ms. Further effects of spike-triggered stimulation conditioning are described in {\it Results}.

\subsubsection{Simulation details}
Spiking network simulations were implemented using Python and MATLAB programming languages with the Mersenne Twister algorithm~\citep{Matsumoto2004} for pseudo-random number generation. A simulation of system~\eqref{model} is performed by discretizing time in bins of width $\d t=0.1$ ms and pseudo-randomly drawing spikes for each neuron $i$ according to the probability $\l_i(t)*\d t$. For each new spike drawn, all affected synapses are updated according to the summation of rule~\eqref{model:plastic} for all temporally-filtered preceding spikes.

For system~\eqref{model_stim}, in the presence of spike-triggered stimulation, every time a spike from neuron 1 is drawn, the spiking probability of every neuron in group $b$ is artificially set to 1, after a delay $d^\dag$. Unless otherwise noted, the parameters used for all spiking simulations are listed in Table~\ref{tparam}.
\begin{table}[h!]
\begin{tabular}{|c|l|l|}
    \hline
      	{\bf Parameter} & {\bf Value} & {\bf Description}\\
	\hline
	$N$ & 60 & network size (simulations)\\
	$p$ & 0.3 & connection probability\\
	$\tau$ & 5 (ms) & synaptic time-constant\\
	$\bar{d}^a$ & 3 (ms) & mean axonal delay\\
	$\sigma^a$ & 1 (ms) & radius of axonal delay distribution\\
	$\bar{d}^d$ & 2 (ms) & mean dendritic delay\\
	$\sigma^d$ & 1 (ms) & radius of dendritic delay distribution\\
	$A^+$ & 30 & STDP potentiation scaling \\
	$A^-$ & 20 & STDP depression scaling\\
	$\tau^+$ & 8.5 (ms) & STDP potentiation time-constant\\
	$\tau^-$ & 17 (ms) & STDP depression time-constant\\
	$J_{min}$ & 0 & synaptic strength lower bound\\
	$J_{max}$ & 0.1 & synaptic strength upper bound\\
	$\gamma$ & 0.1 & STDP weight-dependence exponent\\
	$\n$ & $10^{-8}$ & STDP plasticity rate\\
	$T$ & 2 (sec) & plasticity epoch duration\\
	$d^\dag$ & 10 (ms) & spike-triggered stimulation delay\\	
    \hline
\end{tabular}
\caption{Spiking network simulation parameters}
\label{tparam}
\end{table}

\subsection{Theory}

Our goal is to derive an analytical expression for the dynamics of the synaptic matrix $J(t)$, to allow us to predict the timescale of plastic changes and relative equilibria. Equations~\eqref{model} and~\eqref{model:plastic} show that changes of the synapses stored in $J(t)$, modulated by the spiking activity of the network, depend on the external rates $\nu(t)$ and the synaptic matrix $J(t)$ itself. This leads to a self-consistent relationship between $J(t)$ dynamics and network spiking activity.
Earlier work~\citep{Gilson:2010eh,Gilson:2009iy,Gilson:2009gu,Gilson:2009eq,Gilson:2009ci} describes a framework to effectively decouple these interactions, which we follow and adapt to our needs.

\medskip

The key idea, originally proposed in~\citep{Kempter:1999tn}, is to use a separation of timescales, assuming that synapses are locally constant over time intervals $[t^1,t^2]$, on which the network has stable and stationary spiking statistics. It follows that accumulated increments for the synapse $J_{ij}$ over that interval, denoted $\D J_{ij}=J_{ij}(t^2)-J_{ij}(t^1)$, is given by the sum of all discrete plastic ``steps" due to spike-time pairs $t^s_i,t^s_j \in [t^1,t^2]$ arriving at the $ij$-synapse (ignoring delays for clarity):

\beq
\label{sumJ}
\D J_{ij}=\eta \sum_{t^1\leq t^s_i,t^s_j\leq t^2} W(t^s_j-t^s_i,J_{ij}(t^1)).
\eeq
An equivalent formulation of Equation~\eqref{sumJ}
uses the density of inter-spike-intervals $u=t^s_j-t^s_i$ over the interval $[t^1,t^2]$.
To this end, let $C_{ij}(u)$ be the count of spike pairs separated by $u$ occurring over the interval $[t^1,t^2]$, which is the (non-normalized) cross-correlation between neuron $j$ and neuron $i$.
Then,

\beq
\label{basic}
\D J_{ij}=\eta \int_{-\infty}^{\infty}duC_{ij}(u)W(u,J_{ij}(t^1)).
\eeq
Equation~\eqref{basic}
outlines the basic mechanism governing the evolution of the synaptic matrix $J(t)$ over consecutive time intervals.
The challenge is to express $C_{ij}(u)$ --itself nontrivially dependent on $J(t)$ and external rates $\nu(t)$ -- in a closed form, so that Equation~\eqref{basic} can be iterated by updating $J$ at each step.

\medskip

The authors of \cite{Gilson:2010eh} present detailed descriptions of
this iteration process, and of ways
to estimate the cross-correlations $C_{ij}(u)$.
However, a number of limitations of this original derivation prevent us from using it directly:
(i) The assumption that external rates $\nu_i(t)$ and $\nu_j(t)$ are $\d$-correlated; we require arbitrary cross-correlation functions to fit the model to experiments.
(ii) It is unclear how to include artificial spike-triggered stimulation induced by the BBCI; this introduces non-trivial statistical dependencies between groups and single neurons.

\medskip

In the following, 
still closely following~\cite{Gilson:2010eh}, we present a theoretical derivation that addresses these issues. 
First, we describe parametric constraints that ensure that spiking activity remains stable and does not run away in a self-excitation cascade. Second, we formally describe the timescale separation argument outlined above, followed by estimates for network cross-correlations, both for normal activity and in the presence of spike-triggered stimulation. Finally, we formulate and analyze a dynamical system for averaged synaptic dynamics.

\subsubsection{Network stability}
Here, we find constraints on synaptic weight bounds to ensure that spiking activity does not grow unbounded due to run-away self-excitation.

\medskip
To see how this happens, assume that $J$ is fixed and external rates are constant in time: $\nu(t)\equiv \nu^0$.
It follows that the mean network rates can be approximated by the following expansion:
\beq
\label{rates}
\bar{\l} \sim \nu^0 + J \nu^0 +J^2 \nu_0 +\dots+J^n \nu^0+\mathcal{O}(J^{n+1}).
\eeq
If the terms in this expansion do not decay fast enough, the average activity rates $\bar\l$ quickly diverge. However, it is easy to show that this can be prevented if we require the eigenvalues of $J$ to remain bounded. For more details about precise stability conditions, see~\cite{Gilson:2010eh} and references therein.
For the remainder of this paper, we assume the sufficient, but not necessary condition that all eigenvalues of $J(t)$ remain within the complex unit circle at all times, and adjust the bounds $J_{min}$ and $J_{max}$ accordingly. This means that $J^n \to 0$ as $n\to\infty$, and that for any external rates $\nu(t)$ with stationary statistics, mean rates $\bar{\l}$ are well defined.

\subsubsection{Separation of timescales}
We now present mathematical arguments in support of the simplifying assumption that synaptic matrices remain constant on short time intervals, since synaptic changes occur at slower timescales than spiking activity. 
In turn, this leads to the formulation of well-defined cross-correlation quantities for network activity. The arguments presented here closely follow those originally presented in~\citep{Kempter:1999tn} and
used throughout~\cite{Gilson:2010eh}.

\medskip
The timescale on which impactful plastic changes occur is much longer than the spiking activity timescale, a fact imposed by $\n \ll1$ which implies that only tiny plastic increments can occur for each pair of pre- and post-synaptic spikes.
This allows to effectively separate the spiking dynamics from the synaptic dynamics by assuming that over a reasonably long time period of length $T$ (on the order of a few seconds), synaptic weights $J_{ij}(t)$ are approximately constant ($\approx J_{ij}$). As a result, it is possible to derive dynamic equations for the synaptic connectivity matrix $J$ on the timescale given by $T$.
 Unless otherwise noted, we set $T=2$ seconds for the remainder of this paper.

 As defined in Equation~\eqref{model}, $\l_i(t)$ gives the instantaneous spiking probability of neuron $i$ at time $t$. This probability depends both on the external rate $\nu_i(t)$ and on past spikes from other neurons in the network, which are themselves random processes governed by $\l_j(s)$ for $s<t$.
In what follows, we consider the ensemble average over all these processes to get the expected rate $\la \l_i(t) \ra$. For the sake of convenience, we drop the brackets and write
\beq
\label{model_prob}
\l_i(t)=\nu_i(t)+\sum_j J_{ij} (\l_j*\e^a_{ij})(t)
\eeq
where we define the delayed convolution as
\beqn
(\l_j*\e^a_{ij})(t)\equiv\int_0^tds\l_i(s)\e(t-d^a_{ij}-s).
\eeqn
Equation~\eqref{model_prob} is defined over plasticity epochs $[t-T,t]$ where $J$ is kept constant. At the end of this epoch, we compute the changes in $J$.

The density count of spike pairs separated by $u=t^s_j-t^s_i$ over the epoch $[t-T,t]$ is given by the (non-normalized) cross-correlation between neurons $i$ and $j$: $C_{ij}(u;t)=\la \l_i(s)\l_j(s+u)\ra_t \equiv \int_{t-T}^tds\l_i(s)\l_j(s+u)$. In matrix form, we get
\beq
\label{C_matrix}
\begin{split}
C(u;t)&=\la \l(s)\l^\top(s+u)\ra_t
\end{split}
\eeq
with $\l(t)=(\l_1(t),\dots,\l_N(t))^\top$ and ``$\top$" denoting matrix transposition.
For ``post/pre" neurons ($i$, $j$), the expected number of spike pairs separated by $\D t\in[u_1,u_2]$ in epoch $[t-T,t]$ is given by $\int_{u_1}^{u_2}duC_{ij}(u;t)$.
It follows that the expected incremental change for the synapse $J_{ij}$, after the epoch $[t-T,t]$, is given by

\beq
\label{mean:DJ}
\la \D J_{ij}(t) \ra=\n\int_{-\infty}^{\infty}duC_{ij}(u+d^d_{ij};t)W(u-d^a_{ij},J_{ij}(t-T)).
\eeq
Expression~\eqref{mean:DJ} illustrates that all relevant information needed to predict expected synaptic changes is contained in the network's cross-correlation $C(u;t)$.

\subsubsection{Cross-correlations for normal network activity}
We now derive estimates for cross-correlations of network spiking activity $C(u)$
in terms of $J$ and the statistics of the external driving rates $\nu(t)$,
so that network statistics can be expressed without implicit references to spiking probabilities $\l(t)$.
While a similar quantity is used in~\citep{Gilson:2010eh}, we introduce a new  expansion-based approach that generalizes to arbitrary external statistics.

\begin{figure*}
\includegraphics{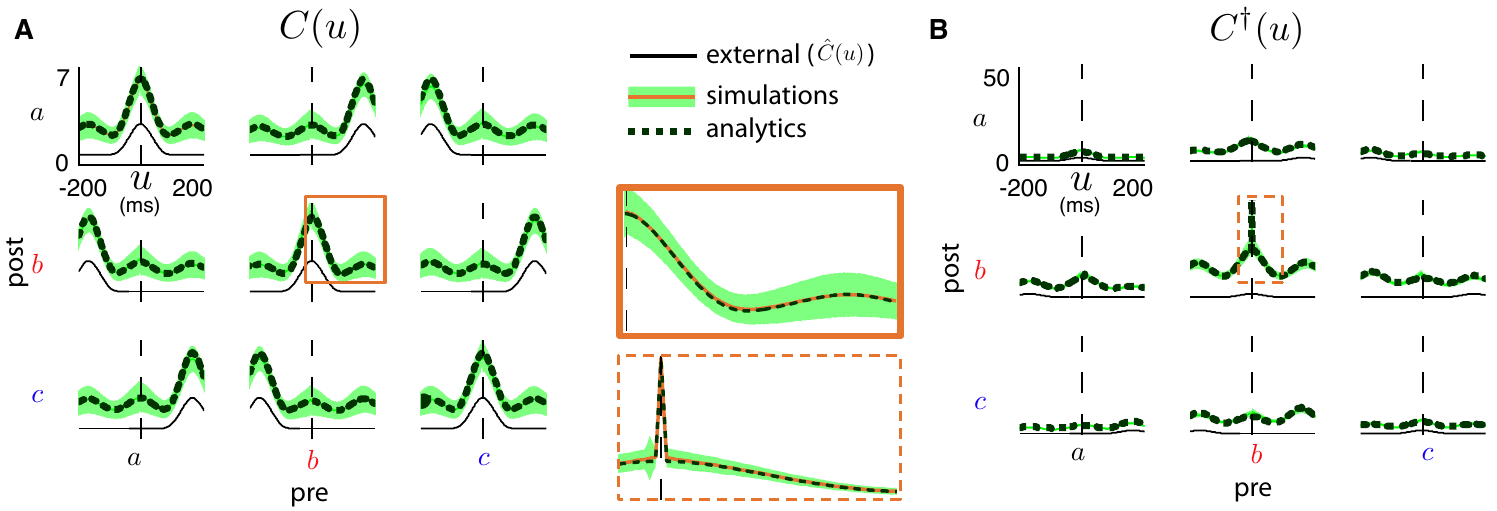}
\caption{Correlation functions averaged over all pairs of connected cells in respective group combinations: external rate correlations $\hat{C}(u)$ (solid black); estimates for network correlations from full spiking network simulations (orange line with shaded green area showing standard deviation of 5 seconds batched estimates) and analytical estimates of order 4 (dashed green).  ({\bf A}) For network under normal activity. ({\bf B}) For network under spike-triggered stimulation ($\dag$). Boxes in the middle show zooms from $C^{(\dag)}(u)$ from A and B.}
\label{fcorrs}
\end{figure*}

\medskip
Consider the (non-normalized) cross-correlation between external rates:
\beqn
\hat{C}(u;t)=\la\nu(s)\nu^\top(s+u)\ra_t
\eeqn
defined as in Equation~\eqref{C_matrix}.
We assume that such ``external" cross-correlations are stationary in time and therefore omit the dependence on $t$: $\hat{C}(u)$.
For our example rates $\nu(t)$ from Figure~\ref{fintro} C, $\hat{C}(u)$ is plotted in Figure~\ref{fcorrs} (thin black line).
How will the network react to external drives with these statistics?

We start by replacing every delay in the network by its mean. As long as the delay distributions are not too wide, this gives accurate results.
This enables us to write \eqref{model_prob} in matrix form:
\beq
\label{model_mat}
\l(t)=\nu(t)+J(\l*\e^a)(t)
\eeq
where $\e^a(t)$ is shorthand for $\e(t-\bar{d}^a)$.
Ideally, one would like to isolate $\l$ from Equation~\eqref{model_mat} but this is difficult because of the convolution with the synaptic kernel $\e$. In~\citep{Gilson:2010eh,Gilson:2009iy,Gilson:2009gu,Gilson:2009eq,Gilson:2009ci}, the authors circumvent this problem by performing calculations in Laplace space, thus transforming the convolution into a product and enabling a self-consistent equation for $\l(t)$. However, this required important assumptions to be made about the shape of external cross-correlations $\hat{C}(u)$ so that inverse transforms remain tractable. In our case, we need $\hat{C}(u)$ to remain as general as possible which makes solving a self-consistent equation exactly challenging.

To circumvent this problem we rely on the stability condition that $J$ must have eigenvalues within the unit circle, and use this to produce expansions that we truncate at higher powers of $J$. We start by substituting Equation~\eqref{model_mat} into the expression~\eqref{C_matrix} for $C(u)$:
\beq
\label{expansion}
\begin{split}
C(u)&=\hat{C}(u)
+\la \nu(s)(\l*\e^a)^\top(s+u)\ra J^\top
+J\la (\l*\e^a)(s)\nu^\top(s+u)\ra
+JC(u)J^\top.
\end{split}
\eeq
To simplify further, we replace the cross-correlation terms involving convolved rates with delayed correlations

\beqn
\la (\l*\e^a)(s)\nu^\top(s+u)\ra  \simeq \la \l(s-\bar{d}^a)\nu^\top(s+u)\ra
 \eeqn
 which is justified for $T$ large enough and $\nu(t)$ fluctuating on longer timescales than $\tau$ ($\e$'s time-constant). In turn, we can further expand this expression to obtain:
 \beq
 \label{cross_cross}
 \begin{split}
  \la\l(s-\bar{d}^a)\nu^\top(s+u)\ra&=\sum_{k=0}^n J^k\hat{C}(u+(k-1)\bar{d}^a)+\mathcal{O}(J^{(n+1)})
 \end{split}
 \eeq
with $\mathcal{O}(J^{(n+1)})\to0$ as $n\to\infty$.  A similar expression can be derived for $\la \nu(s)(\l*\e^a)^\top(s+u)\ra$. Combining Equations~\eqref{expansion} and~\eqref{cross_cross}, we obtain the following truncated, $n^{th}$-order expansion formula:
 \beq
 \label{full_expansion}
 C(u)\simeq\sum_{k=0}^n \sum_{\substack{r,l\geq0\\ r+l=k}} J^r\hat{C}(u+(r-l)\bar{d}^a)J^{\top l}
 \eeq
that provides an algorithm to estimate self-consistent network cross-correlations.
Figure~\ref{fcorrs} A shows $C(u)$ averaged over functional groups, both estimated from spiking simulations and obtained from Equation~\eqref{full_expansion} truncated to fourth order; they agree nicely.

\subsubsection{Cross-correlations for spike-triggered stimulated networks}
We next adapt our expansion-based approach to derive an expression for $C^\dag(u)$: the network's cross-correlation under spike-triggered stimulation. Novel challenges arise since complex statistical dependencies emerge from triggering artificial stimulation on single neuron spikes, and are beyond the scope of the original framework described in~\citep{Gilson:2010eh}.

\medskip
Recall that after a delay $d^\dag$, the entirety of neural group $b$ is artificially stimulated and forced to spike whenever neuron 1 from group $a$ fires a spike (see Equation~\eqref{model_stim}). This is easily implementable in numerical simulations but raises a conceptual difficulty when attempting to transform it into a consistent expression for the densities $\l^\dag(t)$. This is because a particular realization of spike times from neuron $1$ must be fixed across all neurons $i$ in group $b$. This is akin to a Hawkes process~\citep{Hawkes:1971wh} and demands special corrections to be made to $C^\dag(u)$ estimates. We revisit these corrections; for now, we follow the same logic as for Equation~\eqref{model_mat} and write

 \beq
 \label{model_mat_stim}
 \l^\dag(t)=\nu(t)+A\l(t-d^\dag)+J(\l*\e^a)(t)
 \eeq
 where $A_{j1}=1$ for $j\in b=\{N/3+1,...,2N/3\}$. The matrix $A$ adds a copy of the spiking probability from neuron 1 to all neurons in group $b$, after a delay $d^\dag$.

We replace the synaptic filter convolution with the delayed rate, as done in~\eqref{cross_cross}, directly in Equation~\eqref{model_mat_stim}, while keeping in mind these terms will be combined later into cross-correlation terms.
We get
 \beq
 \label{expand_stim}
\l^\dag(t)= \sum_{k=0}^n \sum_{\substack{r,s\geq0\\ r+l=k}}\Pi(A^r,J^l)\nu(t-rd^\dag-l\bar{d}^a)+\mathcal{O}(J^{n})
\eeq
where $\Pi(A^r,J^l)$ denotes the sum of all possible ordered permutations of $r$ times $A$ and $l$ times $J$ (e.g. $\Pi(A^1,J^2)=AJ^2+J^2A+JAJ$). In turn, this leads to the following truncated estimate:
\beq
\label{shitshow}
C^\dag(u)\simeq \sum_{k=0}^n \sum_{\substack{r,l,r',l'\geq0\\ r+l+r'+l'=k}} \Pi(A^r,J^l)\left[\hat{C}(u+(r-r')d^\dag+(l-l')\bar{d}^a) \Pi(A^{\top r'},J^{\top l'}) \right]+R(u;\bar{\l}^\dag)
\eeq
where $R(u;\bar{\l}^\dag)$ is a correction term defined below.
As delays accumulate and since the matrices $A$ and $J$ do not necessarily commute, there is no simple way to concisely express Equation~\eqref{shitshow}. Nevertheless, its simple combinatorial nature enables a relatively straightforward algorithmic implementation. Moreover, while comparing estimates to numerical simulations with current parameters, we find very good agreement with third- or fourth-order estimates.

We still need to account for the fact that a single spike train $S_1(t)\sim\l^\dag_1(t)$ is copied into group $b$ rather than independent realizations sharing the same rate $\l_1^\dag(t)$. Since the rates $\l^\dag(t)$ are treated as ensemble averages ($\langle \l^\dag(t) \rangle$) and assuming independence between disjoint time intervals (as is the case for Poisson processes), it follows that $\langle \l^\dag_i(s)S_1(s+u)\rangle=\langle \l^\dag_i(s)\l_1^\dag(s+u)\rangle$ if $u\neq0$.
To account for the fact that all neurons in group $b$ are driven by a single spike copy, the following modifications are made via the correction term $R(u;\bar{\l}^\dag)$ in Equation~\eqref{shitshow}:

\beq
\label{corrections}
\begin{split}
R_{ij}(u;\bar{\l}^\dag) = & T\bar{\l}^\dag_1\d(u)\\
R_{i1}(u;\bar{\l}^\dag) = & T\bar{\l}^\dag_1\d(u-d^\dag)\\
R_{1i}(u;\bar{\l}^\dag) = & T\bar{\l}^\dag_1\d(u+d^\dag).
\end{split}
\eeq
where $i,j\in b$ and $\bar{\l}^\dag_1$ is defined as in Equation~\eqref{rates} with $\nu^0=\frac{1}{T}\int_{t-T}^t \nu(s) ds$
(the mean over the epoch)
using the same expansion truncation order $n$ used in the cross-correlation estimate.
Figure~\ref{fcorrs} B shows agreement between group-averaged $C^\dag(u)$ derived from Equations~\eqref{expand_stim},~\eqref{corrections} and estimated from spiking numerical simulations.

\subsubsection{Dynamical system for averaged synaptic strengths}
Using the estimates for network cross-correlations with or without BBCI stimulation, $C^{(\dag)}(u)$, we derive a dynamical system for synaptic weights averaged over neural groups. This dynamical system has a temporal resolution of $T$, as it describes the evolution of synapses subject to STDP every $T$-step. Stability arguments are also presented, describing conditions under which one can expect stable, mean synaptic equilibria. Here, we closely follow~\citep{Gilson:2010eh} although stability arguments differ slightly due to the differences in our model presented above.

\medskip
In order to use the correlation estimates to predict synaptic changes, we begin by restricting ourselves to group-averaged quantities. This greatly simplifies calculations and gives remarkably precise estimates for networks where $N$ is large enough. Consider the $3\times3$ matrix $\bar{J}$ whose entries $\bar{J}_{\a\b}$ represent the mean strength of non-zero synapses from a neuron in group $\b$ to a neuron in group $\a$.  Recall that $p$ is the probability that any two cells are connected which means that $\frac{pN}{3}\bar{J}_{\a\b}$ is the mean strength of total synaptic inputs from neurons in group $\b$ to a single neuron in group $\a$.

We follow derivations outlined in the previous section to obtain estimates for the group-averaged network cross-correlations.
For tractability, we explicitly index these estimates by the (averaged) synaptic matrix $\bar J$ and external cross-correlation $\hat C(u)$ used in their derivation: $\bar{C}_{\bar{J},\hat{C}}(u)$ and $\bar{C}_{\bar{J},\hat{C}}^\dag(u)$.
These are now $u$-dependent $3\times 3$ matrices obtained by using
\beqn
\begin{split}
J&=\frac{pN}{3}\bar{J}\\
\hat{C}(u)&=\la [\nu_a,\nu_b,\nu_c](s)[\nu_a,\nu_b,\nu_c]^\top(s+u)\ra\\
A&=\left(\begin{array}{ccc}0 & 0 & 0 \\1 & 0 & 0 \\0 & 0 & 0\end{array}\right)
\end{split}
\eeqn
in equations~\eqref{full_expansion} and \eqref{shitshow}. Furthermore, the correction term $R(u;\bar{\l}^\dag)$ in Equation~\eqref{shitshow} following this group average is replaced by its $3\times3$ averaged counterpart $\bar{R}(u;\bar{\l}^\dag)$ which is zero everywhere except:
\beq
\label{avr_corrections}
\begin{split}
\bar{R}_{bb}(u;\bar{\l}^\dag) = & T\bar{\l}^\dag_a\d(u)\\
\bar{R}_{ba}(u;\bar{\l}^\dag) = & \frac{3}{N}T\bar{\l}^\dag_a\d(u-d^\dag)\\
\bar{R}_{ab}(u;\bar{\l}^\dag) = & \frac{3}{N}T\bar{\l}^\dag_a\d(u+d^\dag).
\end{split}
\eeq
Note that for large $N$, the modification for interactions between groups $a$ and $b$ vanish, but interactions within group $b$ remain independent of $N$, since each cell in $b$ shares the same artificially introduced spikes (see Figure~\ref{fcorrs} B).

\begin{figure*}
\includegraphics{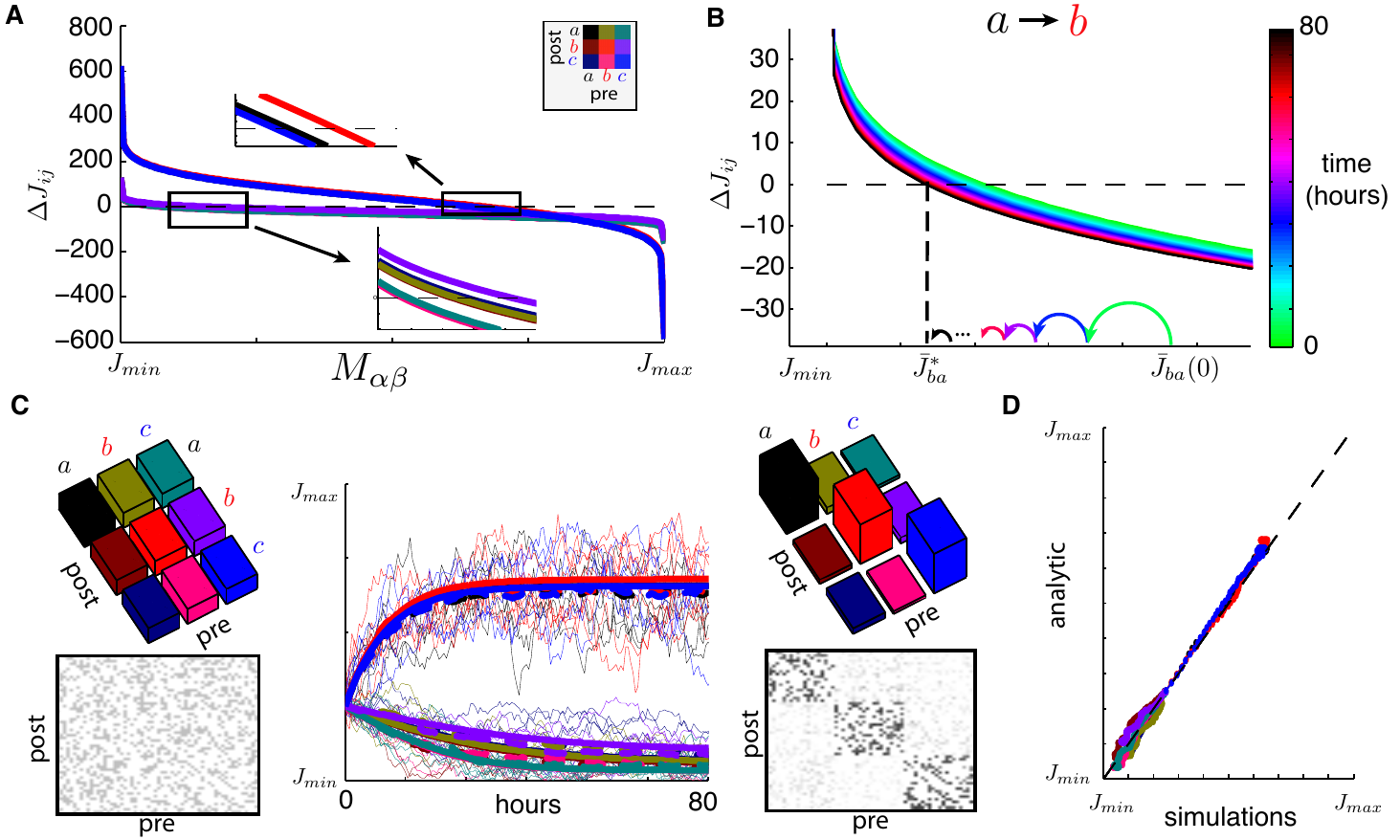}
\caption{({\bf A}) Plot of $F(M_{\a\b},\bar{J}(0),\hat{C}(u))$ (with $\a, \b \in \{ a,b,c \}$) where $J(0)$ is sparse and homogeneous (see C) and $\hat{C}(u)$ is as in Figure~\ref{fcorrs}. Colors represent the pair of pre- and post-synaptic groups $\a\b$ as in the upper left key. ({\bf B}) Plot of $F(M_{ba},\bar{J}_{ba}(t),\hat{C}(u))$ as $\bar{J}(t)$ evolves over time indicated by the color scheme. Curved arrows illustrate the iteration scheme $\bar{J}_{ba}(t+T)=\bar{J}_{ba}(t)+F_{ba}(\bar{J}(t),\bar{J}(t),\hat{C})$ and culminate at the equilibrium $J^*_{ba}$. ({\bf C}) Middle: evolution of synaptic weights over time, color-coded as in A . Thick solid lines show analytical predictions of order 4, thick dotted lines show averages from simulations and thin lines show examples of individual synapses $J_{ij}(t)$. On either sides: plots of $\bar{J}(t)$ and $J(t)$ used for simulations for $t=0$ (left) and $t=80$ hours (right). ({\bf D}) Scatter plot of $\bar{J}_{\a\b}(t)$ and corresponding averages of simulated $J(t)$ over all time points plotted in C. For both C and D, $N=60$, $p=0.3$, $J_{min}=0$, $J_{max}=0.1$, $\n=10^{-8}$ and $J_{ij}(0)=J_{max}/4$.}
\label{fDS}
\end{figure*}

Having a closed expression for $\bar{C}_{\bar{J},\hat{C}}^{(\dag)}(u)$, we define the functional
\beq
\label{function}
F^{(\dag)}(M,\bar{J};\hat{C})\equiv\int_{-\infty}^\infty du \bar{C}_{\bar{J},\hat{C}}^{(\dag)}(u+\bar{d}^d)\circ W\left(u-\bar{d}^a,M\right)
\eeq
where ``$\circ$" designates an element-wise (Hadamard) product and $\bar{C}_{\bar{J},\hat{C}}^{(\dag)}$ refers to the estimated network correlations using mean matrix $\bar{J}$ and $\hat{C}(u)$ with and without spike-triggered stimulation ($\dag$) respectively. From Equation~\eqref{mean:DJ}, $F$ gives the expected average plastic increments for synaptic matrix $M$ (also $3\times 3$), over the epoch $[t-T,t]$.
Figure~\ref{fDS} A shows $F_{\a\b}(M,J;\hat{C})$ for $\bar{J}=\bar{J}(0)$, a random sparse homogeneous matrix as in Figure~\ref{fintro2}, and $\hat{C}(u)$ as in Figure~\ref{fcorrs}.
It follows that $F$ defines a discrete, state-dependent dynamical system on the space of $3 \times 3$ matrices describing the evolution of our network's synapses from one activity epoch to the next:
\beq
\label{DS}
\bar{J}(t+T)=\bar{J}(t)+\n F(\bar{J}(t),\bar{J}(t);\hat{C}).
\eeq
For any initial matrix $\bar{J}(0)$, iterating Equation~\eqref{DS} gives the synaptic evolution of $\bar{J}(t)$ at a temporal resolution $T$.

For stationary external statistics (i.e. $\hat{C}(u)$ does not depend on $t$), a fixed point of the system described by~\eqref{DS} is a matrix $\bar{J}^*$ such that $F(\bar{J}^*,\bar{J}^*;\hat{C})=0$ and defines an averaged synaptic equilibrium matrix. Furthermore, an equilibrium is said to be {\it stable} if nearby initial states ($\bar{J}(0)$) converge toward it. We are interested in the existence and stability of equilibria as they represent homeostatic steady states. However, classical tools of dynamical systems theory such as linearization prove to be inadequate because $F$ is an integro-functional. Nevertheless, some basic geometric observation can provide valuable insights.

\medskip
First, notice that $\bar{C}(u)\geq0$ since it is defined as the non-normalized cross-correlation.
This means that for $M_{ij}$ close to the bounds $\bar{J}_{min}$ and $\bar{J}_{max}$ --where $W(\D t, M_{ij})$ is positive and negative, respectively, for any $\D t$-- $F$ must be positive and negative respectively, for any matrix $\bar{J}$ and any $\hat{C}(u)$.
Moreover, from the definition of $f^{\pm}(M_{ij})$ in Equation~\eqref{J-mult}, it is straightforward
to see that all entries of $F(M,J;\hat{C})$ are monotonically decreasing with respect to each $M_{ij}$ over the interval $[J_{min},J_{max}]$.
Since $F$ is clearly continuous in $M$, it follows from the mean value theorem that there exists a unique $M^*(\bar{J})$ for any $\bar{J}$ such that
\beq
\label{mean_val_th}
F(M^*(\bar{J}),\bar{J},\hat{C})=0.
\eeq
This is illustrated in Figure~\ref{fDS} A where each curve intersects the $\D \bar{J}_{\a\b}=0$ line exactly once.
It follows that a matrix $\bar{J}^*$ is a fixed point of system~\eqref{DS} if and only if it is also a fixed point of $M^*$, i.e., $\bar{J}^*=M^*(\bar{J}^*)$ in Equation~\eqref{mean_val_th}.
It is not clear that there exists a general condition that guarantees that $M^*$ admits a fixed point for finite values of $\n$. Nevertheless we argue that in general, one can expect the dynamics of~\eqref{DS} to converge toward a very small region of matrix space and remain there, as is illustrated in Figure~\ref{fDS} B.

To see why this should be the case, let us assume without loss of generality that at some time $t$, $\bar{J}_{\a\b}(t)>M_{\a\b}^*(\bar{J}(t))$.
By construction, this means that $F_{ij}(\bar{J},\bar{J};\hat{C})<0$ so that $\bar{J}_{\a\b}(t+T)<\bar{J}_{\a\b}(t)$ (i.e. $\bar{J}_{\a\b}$ decreases). Assuming very small steps and by the continuity of $F$, we can expect $M^*$ to also show small changes as $\bar{J}_{\a\b}$ evolves. Thus, the only way the distance between $\bar{J}_{\a\b}$ and $M_{\a\b}^*(\bar{J})$ can grow is if $M_{\a\b}^*(\bar{J}(t))$ also decreases, but this cannot go on forever since $M^*_{\a\b}$ is bounded. It follows that beyond a certain time, $|\bar{J}_{\a\b}(t+T)-M_{\a\b}^*(\bar{J}(t+T))|<|\bar{J}_{\a\b}(t)-M_{\a\b}^*(\bar{J}(t))|$ which implies that $\bar{J}_{\a\b}$ converges toward $M_{\a\b}^*(\bar{J})$.

This contraction argument would be sufficient to show the existence of a fixed point $\bar{J}_{\a\b}^*$ if the function $M_{\a\b}^*$ depended only on the $\a\b$ component of $\bar{J}$, but this is not the case by construction of the functional $F$. Indeed, interactions throughout the network influence the cross-correlation between any two groups $\a$ and $\b$ which in turn, influence expected synaptic changes between these groups.
However, if synapses converge toward their respective fixed points at a comparable rate, then their contributions to $M^*$ all attenuate at similar rates, leading to stable steady states. This is what we observe when we numerically iterate system~\eqref{DS}.

Figure~\ref{fDS} B illustrates this stability property by showing the ``$ba$" coordinate of $F$ throughout the iteration process~\eqref{DS} along with a cartoon representation of iterates $\bar{J}_{ba}(t)$.
We can clearly see that not only does the iterate push $J(t)$ toward a value for which $F=0$, but $F$ also changes in a monotonic manner so that for small enough steps modulated by $\n$, $\bar{J}(t)$ is bound to converge to a stable $\bar{J}^*$. The same is true for each coordinate leading to a global $J^*$.

Figure~\ref{fDS} C shows both spiking network simulations and averaged synaptic dynamics for a system with normal dynamics (i.e. no stimulation) starting from a sparse homogeneous matrix.
Equilibria for averaged synaptic strengths across all functional groups are well captured by the reduced system. Note that while mean synaptic strengths are very stable once an equilibrium is reached, individual synapses are not; they continue to fluctuate as a result of irregular spiking activity and inhomogeneous connectivity.

Moreover, the reduced system~\eqref{DS} also captures the timescale at which group-averaged synapses converge to their respective equilibria since it estimates the expected synaptic changes over plasticity epochs, modulated by the learning rate $\n$.
To better appreciate this point, Figure~\ref{fDS} D shows a scatter plot of $\bar{J}(t)$ from system~\eqref{DS} against spiking simulations for all time points plotted in panel C of the same figure: all points lie close to the diagonal, indicating a tight agreement throughout the convergence process.
We will see in the {\it Results} section that capturing synaptic convergence timescales with our reduced model enables us to calibrate the learning rate $\n$ to fit experimental observations.

Finally, note that to obtain estimates for equilibria alone, a complete iteration procedure is not required. Using the contraction argument presented above, it is sufficient to evaluate $M^*(\bar{J})$ from Equation~\eqref{mean_val_th}
for any initial matrix $\bar{J}$, and repeat the process a few times (i.e. $M^*(M^*(\dots M^*(\bar{J})))$). We find that two or three iterations of this process are enough to converge to the equilibrium $\bar{J}^*$ within machine precision.

\bigskip
In summary, we showed that it is possible to obtain estimates for the network's cross-correlation at a temporal resolution defined by plasticity epochs 
 $[t-T,t]$, for both the normal condition ($C(u;t)$) and the spike-triggered condition ($C^\dag(u;t)$). These estimates can be truncated at any order (i.e. ignoring terms of higher powers of $J$) and depend only on the cross-correlation of external inputs over the same plasticity epoch ($\hat{C}(u;t)$). We find through comparison with numerical simulations that truncating to third or fourth order is sufficient to capture network statistics almost perfectly
as is shown in Figure~\ref{fcorrs}.
These estimates can then be used to create a dynamical system capable of estimating both the stable synaptic equilibria that emerge from spiking dynamics and the convergence timescales leading to them, for both the normal and spike-triggered stimulation paradigms.

\subsection{Experimental Procedure}
In {\it Results}, we use experimentally obtained cross-correlation functions from spiking activity recorded in a monkey implanted with a Utah array in primary MC and performing an isometric 2D target-tracking task. Spikes from multiple electrodes of the array were recorded during task performance, and cross-correlograms were compiled with a resolution of 2 milliseconds. Recorded activity was modulated during task performance. We assume the neurons whose spikes are recorded from the same electrode (using spike sorting) were close enough to be part of the same neural group, in the context of our model. In contrast, neurons recorded from different electrodes, separated by at least 400 $\mu$m, are considered as part of different groups.

\section{Acknowledgments}
The authors thank Alain Vinet for insightful discussions, and Andrew Bogaard and Andrew Jackson for experimental data. 
This work was funded in part by the Washington Research Foundation (WRF) UW Institute for Neuroengineering ({\it GL}), the National Science Foundation (NSF) Center for Sensorimotor Neural Engineering at the University of Washington, NIH grant NS12542 ({\it EF}), an Allen Distinguished Investigator grant ({\it ALF}), and CIHR grant MOP-84454 and HFSP RGP0049/2009-C ({\it JFK} and {\it NIK}).


\end{document}